\documentclass[conference]{IEEEtran}
\IEEEoverridecommandlockouts
\usepackage{cite}
\usepackage{amsmath,amssymb,amsfonts}
\usepackage{graphicx}
\usepackage{textcomp}
\usepackage{xcolor}

\usepackage{amsmath}
\usepackage{amsfonts}
\usepackage{amssymb}
\usepackage{multirow}
\usepackage{algorithm}
\usepackage{algpseudocode}
\usepackage{graphicx}
\usepackage{url}
\usepackage{amsthm}
\usepackage{subcaption}

\graphicspath{{figures/}}

\newtheorem{theorem}{Theorem}[section]
\newtheorem{lemma}{Lemma}[section]
\newtheorem{definition}{Definition}[section]

\newtheorem{example}{Example}[section]

\usepackage{color,soul}
\newcommand{\revA}[1]{\textcolor{black}{#1}}

\usepackage{tikz}

\usepackage{environ}
\usepackage{setspace} 

\newenvironment{alignSmall}{\nobreak\small\noindent\align}{\endalign}
\newenvironment{alignFootnotesize}{\nobreak\footnotesize\noindent\align}{\endalign}
\newenvironment{alignScriptsize}{\nobreak\scriptsize\noindent\align}{\endalign}

\NewEnviron{myAlignS}{%
	\begin{singlespace}
		\vspace{-3pt}
		\begin{alignSmall}
			\BODY
		\end{alignSmall}
		\vspace{-5pt}
	\end{singlespace}
}

\NewEnviron{myAlignSS}{%
	\begin{singlespace}
		\vspace{-3pt}
		\begin{alignFootnotesize}
			\BODY
		\end{alignFootnotesize}
		\vspace{-5pt}
	\end{singlespace}
}

\NewEnviron{myAlignSSS}{%
	\begin{singlespace}
		\vspace{-5pt}
		\begin{alignScriptsize}
			\BODY
		\end{alignScriptsize}
		\vspace{-5pt}
	\end{singlespace}
}

\begin{document}

\title{PriSTE: From Location Privacy to Spatiotemporal Event Privacy}


\author{\IEEEauthorblockN{Yang Cao\IEEEauthorrefmark{1}\IEEEauthorrefmark{5}\IEEEauthorrefmark{6}\thanks{Yang and Yonghui contributed equally to this work.}, Yonghui Xiao\IEEEauthorrefmark{3}\IEEEauthorrefmark{7}, Li Xiong\IEEEauthorrefmark{2}\IEEEauthorrefmark{5}, Liquan Bai\IEEEauthorrefmark{4}\IEEEauthorrefmark{5}
\IEEEauthorblockA{\IEEEauthorrefmark{5}\textit{Emory University}, Atlanta, U.S.A \\}
\IEEEauthorblockA{\IEEEauthorrefmark{6}\textit{Kyoto University}, Kyoto, Japan \\}
\IEEEauthorblockA{\IEEEauthorrefmark{7}\textit{Google Inc.}, Mountain View, U.S.A \\}
Email: \IEEEauthorrefmark{1}yang@i.kyoto-u.ac.jp \IEEEauthorrefmark{2}{lxiong}@emory.edu, \IEEEauthorrefmark{3}yhandxiao@gmail.com, \IEEEauthorrefmark{4}bailiquan@gmail.com}
}

\maketitle

\thispagestyle{plain}
\pagestyle{plain}

\begin{abstract}
Location privacy-preserving mechanisms (LPPMs) have been extensively studied for protecting a user's location at each time point or a sequence of locations with different timestamps (i.e., a trajectory).
We argue that existing LPPMs are not capable of protecting the sensitive information in user's spatiotemporal activities, such as ``visited hospital  in the last week" or ``regularly commuting between Address 1 and Address 2 every morning and afternoon" (it is easy to infer that Addresses 1 and 2 may be home and office).
We define such privacy as \textit{Spatiotemporal Event Privacy}, which can be formalized as Boolean expressions between location and time predicates.
To understand how much spatiotemporal event privacy that existing LPPMs can provide, we first formally define spatiotemporal event privacy by extending the notion of differential privacy, and then provide a framework for calculating the spatiotemporal event privacy  loss of a given LPPM under attackers who have knowledge of user's mobility pattern.
We also show a case study of utilizing our framework  to convert the state-of-the-art mechanism for location privacy, i.e., Planner Laplace Mechanism for Geo-indistinguishability, into one protecting spatiotemporal event privacy.
Our experiments on real-life  and synthetic data verified that the proposed method is effective and efficient.


\end{abstract}

\section{Introduction}
The continued advances and usage of smartphones and GPS-enabled devices have provided tremendous opportunities for Location-Based Service (LBS), such as Google Maps, Facebook Places and Swarm.
In the location-based services, mobile users have to share their locations or trajectories with the service providers in order to issue snapshot or continuous queries, for example, ``where is the nearest gas station'' or ``continuously report the taxis within one mile of my location''.
It has raised  privacy concerns as users' digital trace can be used to infer sensitive information, such as home and work place,  religious places and sexual inclinations\cite{golle_anonymity_2009}\cite{recabarren_what_2017}\cite{argyros_evaluating_2017}.

A large number of studies (see surveys\cite{krumm_survey_2009}\cite{wernke_classification_2014}\cite{chatzikokolakis_methods_2017}\cite{liu_location_2018}) have explored how to protect user's location privacy from different aspects:  privacy goals, adversarial models, location privacy metrics, and Location Privacy Preserving Mechanisms (LPPMs).
\textit{Privacy goals} indicate what should be protected or what are the secrets \revA{(e.g., a  single location or a trajectory)}; \textit{adversarial models} make assumptions about the adversaries; \textit{location privacy metrics} formally define the quantitative method for the privacy goal \revA{(e.g., Geo-indistinguishability \cite{andres_geo-indistinguishability:_2013} or $\delta$-location set privacy \cite{xiao_protecting_2015})};  LPPMs study how to achieve  a specified privacy metric.

\revA{We argue that  existing LPPMs may not be able to fully protect users' sensitive information in their spatiotemporal activities because  their privacy goal is focused on protection of either a single location or a trajectory. 
 A user's location data can be represented by tuple (we consider the single user setting in this paper), i.e., $ < $\textit{location, time}$ > $, which corresponds to information about ``where'' and ``when'' in user's real-world activities.
 Hence,  the privacy goals in literature can be categorized into protecting a single \textit{position} or a \textit{trajectory}.
 Many LPPMs are proposed for these goals based on different privacy metrics.
For example, Gruteser et al. \cite{gruteser_anonymous_2003} designed a spatiotemporal cloaking mechanism satisfying k-anonymity to protect movement trajectories of users; 
Andr{\'e}s et al. \cite{andres_geo-indistinguishability:_2013} proposed Planar Laplace mechanism \cite{andres_geo-indistinguishability:_2013} achieving Geo-indistinguishability to protect single locations; Xiao and Xiong \cite{xiao_protecting_2015} designed  Planar Isotropic Mechanism for $\delta$-location set privacy to protect each location in a trajectory.
}

\begin{figure}[h]
 \centering
 \includegraphics[width=7cm]{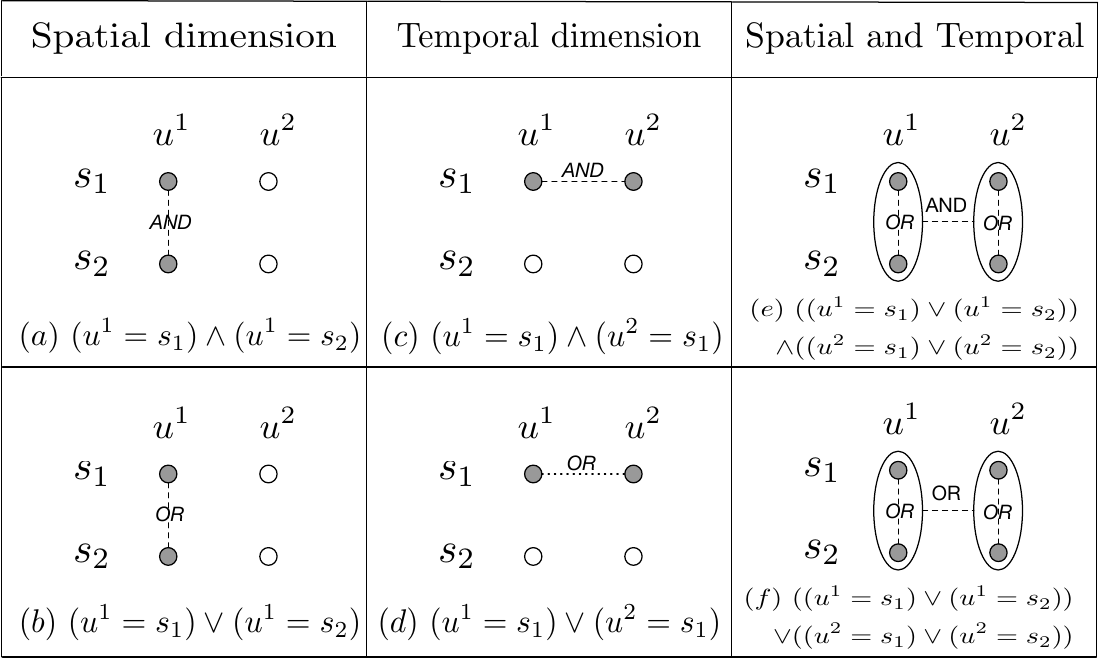}
 \caption{Examples of spatiotemporal events. $ s_1$ and $s_2 $ are two locations on the map $\mathbb{S} $. $ u^1 $ and $ u^2 $ are two variables about user's possible locations at time point $ 1 $ and time point $ 2 $, respectively. Event (a) is always false since a user cannot be at two different locations at the same time. Event (b) means that the secret is a sensitive area including locations $ \{s_1,s_2\} $. Event (c) represents a sensitive trajectory $ s_1 \rightarrow s_1 $. Event (d) denotes that the secret is the visit to $ s_1 $ at time point $ 1 $ \textit{or}  $  2$. Event (e) depicts the secret as a type of trajectory \textbf{pattern}, i.e., the user may stay at two sensitive areas successively. Event (f) indicates the secret as user's {\textbf{presence}} in sensitive area $ \{s_1,s_2\} $ at  either time point $ 1 $ or  $ 2 $.}
 \label{fig:stevent}
 \end{figure}

\revA{However, the privacy goals  in the literature of location privacy only attempt to confuse adversaries about either the user's  exact  location or trajectory, and the two types of privacy goals cannot cover all cases of complex combination of spatial and temporal information (as shown in Fig.\ref{fig:stevent}), which we refer to as \textit{spatiotemporal events} in this paper.}
 Examples of spatiotemporal event include ``visited hospital  in the last week" (i.e., the hospital visit may happen once or multiple times at any time in last week) and ``regularly commuting between Address 1 and Address 2 every morning and every afternoon" (these periodic spatiotemporal events may happen every day).
 
 \revA{We show six cases of the Boolean expression between location and time predicates in Fig.\ref{fig:stevent}.
 It turns out that protecting a single location or a trajectory are only two cases among possible privacy goals in protecting a user's spatial and temporal information.
 Let  $ u^t $  be a user's position at time $ t $, and $ s_i\in \mathbb{S}, i\in[1,m]$ be one of all $ m $ locations on the map.
 As shown in Fig.\ref{fig:stevent},  the element of a user's secrets in her spatiotemporal activities can be represented by a predicate $ u^t=s_i $ (the value can be either \textit{true} or \textit{false}).}
Then, a \textit{spatiotemporal event} can be defined as a Boolean expression by combining different predicates over spatial and/or temporal dimensions (a predicate alone also can be a spatiotemporal event).
As shown in Fig.\ref{fig:stevent}, the events representing a sensitive location/area and a trajectory, which are the main focuses in previous studies of location privacy, are only two cases (i.e., (b) and (c)) in the six enumerated examples.
\revA{Even if each location or a trajectory is protected, it is not clear whether or not adversary is  able to infer  the value of a sensitive spatiotemporal event. 
Protecting the privacy of spatiotemporal events has not been studied in literature.}
 
\vspace{-3pt}

\revA{Although an LPPM protecting a single location or a trajectory (i.e., Fig.\ref{fig:stevent}(b) or Fig.\ref{fig:stevent}(c)) ensures user's location privacy, it is not clear whether such mechanisms also provide 
a certain level of spatiotemporal event privacy (such as  Fig.\ref{fig:stevent}(e) or Fig.\ref{fig:stevent}(f)).
In this paper, we attempt to formalize  the new privacy goal of spatiotemporal event privacy and quantify how much spatiotemporal event privacy that an existing LPPM can provide.}
Towards this goal,  in addition to the lack of definition and privacy metrics for spatiotemporal event, a challenge  is the computational complexity for quantifying the privacy loss of a given LPPM w.r.t a specified spatiotemporal event. 
For example, given a complex spatiotemporal event, i.e., a Boolean expression, checking its value (true or false)  requires enumeration of all possible values of the predicates in the Boolean expression, which can be exponential to the number of predicates.

\vspace{2mm}\noindent
{\bf Contributions.}
In this paper,  we study how to  protect spatiotemporal event privacy for the first time. 
Our contributions  are summarized as follows.

First, we investigate the privacy goal and privacy metric for spatiotemporal event privacy.
We formally define the new type of privacy goal, i.e., spatiotemporal events, as Boolean expressions of a set of (location, time) predicates, and propose a privacy metric for protecting spatiotemporal events by extending  the notion of differential privacy which we call \textit{$ \epsilon $-spatiotemporal event privacy}.
We also explore the difference between the metrics of location privacy and spatiotemporal event privacy.
It turns out that, although the privacy goal of spatiotemporal event privacy is more general than location privacy, the privacy metrics can be orthogonal.
Hence, it would be preferable that an LPPM achieving a location privacy metric such as Geo-indistinguishability can also satisfy $ \epsilon $-spatiotemporal event privacy w.r.t. user-specified events.
Location privacy  provides general protection against unknown risks, while spatiotemporal event privacy guarantees flexible and customizable protection which may not be provided by the existing LPPMs.

Second, we develop a method for quantifying how much $ \epsilon $-spatiotemporal event privacy  a given LPPM can provide.
In this paper, we model  an LPPM as an emission matrix that takes user's true position as input and outputs a perturbed location.
As we mentioned previously, one of the challenges in quantifying the probability of a  spatiotemporal event is that the computational complexity may be exponentially increasing with the number of its predicates.
We develop a two-possible-world method to quantify spatiotemporal event privacy with  linear complexity to the number of predicates.

Third, based on our quantification method, we propose a framework, i.e, PriSTE (\underline{Pri}vate \underline{S}patio-\underline{T}emporal \underline{E}vent), which converts a mechanism for location privacy  into one for spatiotemporal event privacy.
 We demonstrate the effectiveness of our framework using \revA{two case studies using Geo-indistinguishability \cite{andres_geo-indistinguishability:_2013} and $\delta$-location set privacy \cite{xiao_protecting_2015}, which are the state-of-the-art location privacy metrics.}




Finally, we implement and evaluate our algorithms on both synthetic and real-world datasets evaluating its feasibility, efficiency, and the impact of various parameters.


\vspace{-5pt}
\section{Problem Setting and Definitions}

\subsection{Problem Setting}
\label{subsec:problem-statement}
We study how to protect spatiotemporal event in a single user setting.
Consider a user who is sharing her location sequence with a location-based service provider.  
\revA{Since a system with a trusted server is vulnerable to single point of attack, the user does not want to share her sensitive information with the server; instead, she uses a local LPPM that guarantees location privacy at each timestamp. }
We denote a moving user's true locations as {\footnotesize $\{u^1,u^2,\cdots,u^T\}$}.
The LPPM blurs user's true location $  u^t$ to a perturbed one $  o_t$ that satisfies a privacy metric such as \textit{geo-indistinguishability}\cite{andres_geo-indistinguishability:_2013} or \textit{$\delta$-location set privacy} \cite{xiao_protecting_2015}.
Hence, the LPPM can be considered as an emission matrix that takes user's true location as input and outputs a perturbed one.
The major notations in this paper are summarized in Table \ref{tbl-denotation}.
\renewcommand{\arraystretch}{1.2}
\begin{table}[!htbp]
	\scriptsize
	\begin{tabular}{|p{20pt}|p{200pt}|}
		\hline
		$\textbf{s}$ & a region consists of some states, $\textbf{s} \in \{0,1\}^{m\times 1}$ \\\hline
		$t$ &one timestamp in $\{1,2,\cdots, T\}$\\\hline
		\revA{$\textbf{S}$} & \revA{a set of  regions $\textbf{s}$}  \\\hline
		\revA{$\textbf{T}$} & \revA{a set of timestamps } \\\hline
		$u^t$ & a user's true location at time $t$\\ \hline
		$o^t$ & a user's perturbed location at time $t$\\ \hline
		$\textsc{Event}$ & a spatiotemporal event in time $\{start,\cdots,end\}$\\\hline
	\revA{	$\tilde{\textbf{p}}_{o_t}$} & \revA{a vector of emission probabilities given the observation $o_t$}.\\\hline
		\revA{$\tilde{\textbf{p}}_{o_t}^\textbf{D}$} &  \revA{a diagonal matrix with the vector $\tilde{\textbf{p}}_{o_t}$ on the diagonal.} \\\hline
	\end{tabular}
	\caption{Notations}
	\label{tbl-denotation}
\end{table}

\subsection{Spatiotemporal Events}
Spatiotemporal events can represent user's secrets in their real-world activities, such as ``visited hospital in the last week" or ``commuting between Address 1 and Address 2 every morning and afternoon".
Let $\mathbb{S}=\{s_1,s_2,\cdots, s_m\}$  be the domain of space, where $m$ is the number of all locations and $ s_i $ is one location (we use \textit{state} interchangeably) on the map.
A user's trajectory consists of a set of $\{u,t\}$ denoting the user's location at timestamp $t$ in $\{1,2,\cdots,T\}$.
Each pair of location and time can be represented by a predicate.
For example, the pair $\{u^1,s_{3}\}$ can be denoted by a predicate $u^{1}=s_{3}$.
If the user is in location $s_{3}$ at timestamp $1$, then the ground truth of the predicate is true. 
A spatiotemporal event is defined as a Boolean expression of the (location, time) predicates using the AND, OR, NOT operators, denoted by $\land$, $\lor$, $\neg$ respectively.

\begin{table*}[!htbp]
	\scriptsize
	\centering
	\begin{tabular}{|c|c|c|}
		\hline
		{\centering {\bf \textsc{Event}}}&{\centering {\bf Boolean Expression} }&{\centering {\bf Interpretation} }\\
		\hline
		single location&$u^{t}=s_{i}$& the location at timestamp $t$ is $s_{i}$ \\\hline
		$\textsc{Presence}$ at $s_{i}$ during $T$  &$(u^{1}=s_{i})\lor (u^{2}=s_{i})\lor \cdots \lor (u^{t}=s_{i})$& appears at location $s_{i}$ during time $\{1,2,\cdots,T\}$ \\\hline
		$\textsc{Presence}$ at $\textbf{s}$ and $t$  &$(u^{t}=s_{i})\lor (u^{t}=s_{j})\lor \cdots, \lor (u^{t}=s_{k})$ &  appears in region $\textbf{s}:\{s_{i},\cdots,s_{k}\}$ at timestamp $t$\\\hline
		single trajectory&$(u^{1}=s_{i})\land (u^{2}=s_{j})\land \cdots,\land (u^{n}=s_{k})$&a trajectory of locations during a time period\\\hline
		$\textsc{Pattern}$ of trajectories & \begin{tabular}[l]{@{}c@{}}$((u^{1}=s_{i})\lor (u^{1}=s_{j})\lor\cdots,\lor (u^{1}=s_{k}))\land\cdots $\\$\land((u^{n}=s_{\revA{l}})\lor (u^{n}=s_{\revA{m}})\lor\cdots, (u^{n}=s_{\revA{n}}))$\end{tabular}& a \textsc{Pattern} of trajectories\\\hline
	\end{tabular}
	\caption{{\small $\textsc{Events}$ of Boolean operations on the (location, time) predicates}}
	\label{tbl-events}
	\vspace{-18pt}
\end{table*}

\begin{definition}[$\textsc{Event}$]
	A spatiotemporal event, denoted by $\textsc{Event}$, is a set of (location, time) predicates, i.e. $u^{t}=s_{i}$, under the Boolean operations.
\end{definition}
%
Using Boolean logic to define spatiotemporal events enable  users to customize their privacy preference in diverse real-world activities.
Table \ref{tbl-events} shows some representative examples of $\textsc{Event}$. 
If a user is in a state $s_i$ at timestamp $t$, then $u^t=s_i$. 
If the user is in a region of 
$\{s_{i},s_{j},\cdots,s_{k}\}$ at timestamp $t$, then $(u^{t}=s_{i})\lor (u^{t}=s_{j})\lor \cdots,\lor (u^{t}=s_{k})$ holds. 
If the trajectory of the user is $\{s_{i},s_{j},\cdots,s_{k}\}$ over timestamps $1$ to $T$, then $(u^{1}=s_{i})\land (u^{2}=s_{j})\land \cdots,\land (u^{t}=s_{k})$ holds. 
%
Based on the Boolean operations, complicated spatiotemporal events can be defined as follows.

\begin{figure}[t]
	\centering
	\includegraphics[width=5cm]{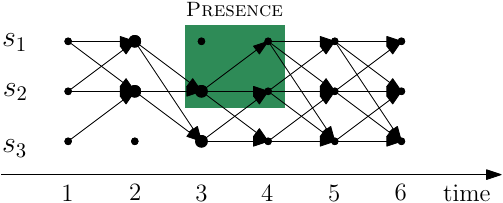}
	\caption{{\small $\textsc{Presence}$: the user appears in regions of $\{s_{1}, s_{2}\}$ during timestamps $3$ and $4$; \textsc{Pattern}: the trajectories all go through $\{s_{1},s_{2}\}$ at timestamp $2$ and $\{s_{2},s_{3}\}$ at timestamp $3$.}}
	\label{Figure-MM-example1}
\end{figure}

\noindent{\bf \textsc{Presence}.}
When the secret is whether or not a user visited a sensitive area (e.g., medical	facilities) in a given time period,
we can use \textsc{Presence} to represent such secret.
A \textsc{Presence} event holds if a user appears in a region during some time. 
In the simplest case, the region consists of one location, and time period consists of one timestamp, then it becomes one single location shown in Table \ref{tbl-events}.
Hence, \textsc{Presence} is a generalization of secrets about single locations.
To denote a region, which is a set of locations, we use a vector $\textbf{s}\in \{0,1\}^{m\times 1}$ where the $i$th element is $1$ if  the region contains $s_i$. The time period is denoted by $\textbf{T}$ as a set of timestamps.
\begin{definition}[\textsc{Presence}]
	\label{def-presence}
	Given a set of regions $ \textbf{S} $ and a time period $\textbf{T}$, if a user appears in $\textbf{s}$ at any timestamp $t\in\textbf{T}$, 
	then it is a presence event, denoted by
	$\textsc{Presence}(\textbf{S},\textbf{T})$.
\end{definition}
\begin{example}[\textsc{Presence}]
	Fig.\ref{Figure-MM-example1} shows a map of $\mathbb{S}=\{s_{1},s_{2},s_{3}\}$. 
	The shaded region shows a \textsc{Pattern} event that the user appears in a region of $s_{1}$ or $s_{2}$ during timestamps $3$ and $4$. 
	The lines indicate possible trajectories.
	As long as user's true trajectory passes through the shaded region, the ground truth of the event is true.
	For this event, the region $\textbf{s}=[1,1,0]^{\intercal}$ denoting the states $s_{1}$ and $s_{2}$; the time period $\textbf{T}=\{3,4\}$ denoting timestamp $3$ and $4$. 
	The \textsc{Presence} event is expressed as $(u^{3}=s_{1})\lor(u^{3}=s_{2})\lor(u^{4}=s_{1})\lor(u^{4}=s_{2})$.
\end{example}
\vspace{-5pt}

%
\noindent{\bf \textsc{Pattern}.}
When the secret is whether or not a user visited multiple sensitive areas successively (e.g., a love hotel and then home),
we can use \textsc{Presence} to represent such secret.
In a simple case, the regions consist of single locations at a set of timestamps, then it becomes single trajectory shown in Table \ref{tbl-events}.
Hence, \textsc{Pattern} is a generalization of secrets about user's trajectories.
We define \textsc{Pattern} as follows where the set of regions $[\textbf{s}_1,\textbf{s}_2,\cdots,\textbf{s}_n]$ is denoted by $\textbf{S}$. 
\begin{definition}[\textsc{Pattern}]
	\label{def-pattern}
	Given a sequence of regions $\textbf{S}=\{\textbf{s}_1,\textbf{s}_2,\cdots,\textbf{s}_n\}$ and 
	a time period $\textbf{T}$ where $\textbf{s}_{i}\in \{0,1\}^{m\times 1}$, if a user appears in $\{\textbf{s}_1,\textbf{s}_2,\cdots,\textbf{s}_n\}$ sequentially during $\textbf{T}$, 
	then it is a pattern event, denoted by $\textsc{Pattern}(\textbf{S},\textbf{T})$.
\end{definition}
\begin{example}[\textsc{Pattern}]
	Fig.\ref{Figure-MM-example1} shows a set of trajectories with a \textsc{Pattern} that all trajectories go through {\footnotesize $\{s_{1},s_{2}\}$} at timestamp $2$ and {\footnotesize $\{s_{2},s_{3}\}$} at timestamp $3$. For this event, 
	the region at timestamp $2$ is {\footnotesize $\textbf{s}_{2}=[1,1,0]^{\intercal}$} denoting $s_{1}$ and $2_{2}$;
	the region at timestamp $3$ is {\footnotesize $\textbf{s}_{3}=[0,1,1]^{\intercal}$} denoting $s_{2}$ and $s_{3}$.
	The \textsc{Pattern} event is expressed as {\footnotesize $((u^{2}=s_{1})\lor(u^{2}=s_{2}))\land((u^{3}=s_{2})\lor(u^{3}=s_{3}))$}.
\end{example}


From the above definitions, we can see that, in terms of privacy goal, spatiotemporal event privacy is a generalization of location privacy.
In this paper, we focus on the two representative  events defined above, i.e., \textsc{Presence} and \textsc{Pattern}, which are the two most complicated events in examples of Fig.\ref{fig:stevent}. 
We note that $\textsc{Presence}$ and $\textsc{Pattern}$  include the cases when the time $\textbf{T}$ is not consecutive.
For simplicity, we assume that the events are defined in consecutive time and use $ start $ and $ end $ to denote the start point and end point \revA{of the defined spatiotemporal event}.
Users can customize one or multiple  spatiotemporal events to be protected.

Protecting spatiotemporal events requires a \textit{local} mechanism that is able to localize the information about spatiotemporal event. 
We need a formal privacy metric to preserve user's \textit{plausible deniability} about the truth of her specified spatiotemporal events, so that even when adversaries happen to infer the specified spatiotemporal event, the user has plausible deniability.
We propose such a privacy metric for spatiotemporal event privacy in the next section.

\vspace{-8pt}
\subsection{$ \epsilon $-Spatiotemporal Event Privacy}

Inspired by the definition of differential privacy\cite{dwork_differential_2008}, we define $ \epsilon $-Spatiotemporal Event Privacy as follows.
\begin{definition}[$\epsilon$-Spatiotemporal Event Privacy]
	\label{def-eps-delta-DP}
	A mechanism preserves $\epsilon$-Spatiotemporal Event Privacy for a spatiotemporal \textsc{Event} if at any timestamp $t$ in $\{1,2,\cdots,T\}$ given any observations $\{o_1,o_2,\cdots,o_T\}$,
	\begin{myAlignSS}
	\label{eqn-eps-delta-DP1}
	\Pr(o_1,o_{2},\cdots,o_t|\textsc{Event}) \leq e^{\epsilon}\Pr(o_1,o_{2},\cdots,o_t| \lnot\textsc{Event})
	\end{myAlignSS}
\vspace{-5pt}
	\vspace{-5pt}
	where  {\textsc{Event}} is a logic variable about the defined spatiotemporal event and  $\lnot\textsc{Event}$ denotes the negation of $\textsc{Event}$. \revA{{\small $ \Pr(o_1,o_{2},\cdots,o_t|\textsc{Event}) $} denotes the probability of the observations $ o_1,o_{2},\cdots,o_t $ given the value of \textsc{Event}. }
\end{definition}

\revA{There are two major benefits of adopting such ``DP-like'' privacy metric.
First, it provides a well-defined semantics for spatiotemporal event privacy. 
Similar to differential privacy that requires the indistinguishability between any two neighboring databases\cite{dwork_differential_2008}, $\epsilon$-Spatiotemporal Event Privacy requires the indistinguishability regarding whether  the $\textsc{Event}$ is true or false given any observations.
Another benefit is that, similar to differential privacy whose privacy  guarantee is independent of the prior probability of a given databases,  the privacy provided by $\epsilon$-Spatiotemporal Event Privacy  is independent of the prior probability of the given spatiotemporal event\footnote{\revA{In our approach,  we first compute the prior probability of the given event with a specific initial distribution $ \pi $, and then we make sure that the privacy is guaranteed w.r.t. any  $ \pi $ in Section \ref{sec:priste}.}}.}

\begin{figure}[t]
	\centering
	\includegraphics[width=9cm]{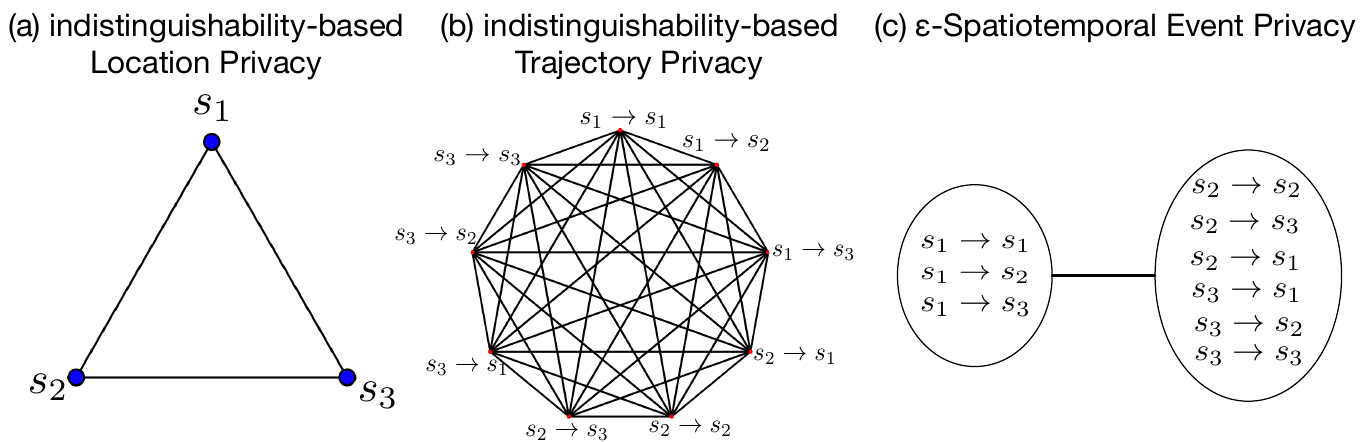}
	\caption{Illustration of  indistinguishability-based privacy metrics for distinct privacy goals when {\footnotesize $ {\mathbb{S}}=\{s_1,s_2,s_3\}$} and {\footnotesize $  T=2$}.}
	\label{fig:secrets_pairs}
\end{figure}

Although the \textit{privacy goal} of spatiotemporal event privacy can be considered as a generalization of location privacy, we note that it may not be true in terms of \textit{privacy metrics}.
We illustrate the indistinguishability-based privacy metrics for the three privacy goals in Fig.\ref{fig:secrets_pairs}, where the lines connecting two secrets indicate the requirements of indistinguishability between the corresponding two possible values of the secrets.

As shown in Fig.\ref{fig:secrets_pairs} (a), indistinguishability-based location privacy metrics (such as geo-indistinguishability\cite{andres_geo-indistinguishability:_2013}) require indistinguishability  between each pair of locations.
Indistinguishability-based trajectory privacy metrics \cite{xiao_protecting_2015} \cite{chatzikokolakis_predictive_2014}\cite{theodorakopoulos_prolonging_2014} requires indistinguishability between each pair of possible trajectories as shown in Fig.\ref{fig:secrets_pairs} (b).
Whereas, $ \epsilon $-spatiotemporal event privacy only requires  indistinguishability between the defined event and its negation.
For example, if the spatiotemporal event is defined as $\textsc{Pattern}(\textbf{S},\textbf{T})$ where $  \textbf{S}=\{\textbf{s}_1, \textbf{s}_2\}, \textbf{s}_1=\{s_1\}, \textbf{s}_2=\{s_1,s_2,s_3\}$ and $ \textbf{T}=\{1,2\} $ (i.e., a trajectory passes through $ s_1 $ and  $ \{s_1,s_2,s_3\} $) successively, then it does not require the indistinguishability between $ \{s_2 \rightarrow s_3 \} $ and $ \{s_3 \rightarrow s_2 \} $ as shown in Fig.\ref{fig:secrets_pairs}(c).
Such privacy requirement makes sense; for example, $ s_1 $ can be a ``love hotel'', $ s_2 $ is ``home'',  and $ s_3 $ is ``office''.
Such spatiotemporal event privacy may not be protected by LPPMs.

While we can define simple events such as a location or trajectory, the corresponding $ \epsilon $-spatiotemporal event privacy does not imply the indistinguishability-based location privacy or trajectory privacy.
For example, even if a user specifies all possible trajectories as her  requirements  for $ \epsilon $-spatiotemporal event privacy, it only ensures the indistinguishability between ``one trajectory'' and ``not this trajectory'', but  no guarantee on the indistinguishability between any two trajectories.
Hence, the privacy guarantee of $ \epsilon $-spatiotemporal event privacy is orthogonal to geo-indistinguishability.

\revA{Event-level or user-level differential privacy \cite{dwork_differential_2010} is different from spatiotemporal event privacy.
In such definitions, an event  means  user's   information at a single time point (different to the definition of spatiotemporal event in this paper); hence, event-level differential privacy preserves the indistinguishability between any two possible values of a single event.
User-level differential privacy preserves the  indistinguishability between any two possible  sequence of events.
These privacy metrics are for time-series data and statistical queries.
Despite the difference of the settings,  the indistinguishability considered in event-level/user-level differential privacy are similar to the above location privacy and trajectory privacy, which are shown in  Fig.\ref{fig:secrets_pairs}(a)(b).
Whereas the indistinguishability considered in spatiotemporal event privacy are shown in Fig.\ref{fig:secrets_pairs}(c).
}

\revA{Location privacy  provides general protection against unknown risks when sharing location with the third parties, while spatiotemporal event privacy guarantees flexible and customizable protection which may prevent against profiling attacks such as inferring user's trajectory pattern (location privacy cannot provide such protection).
Therefore, it would be preferable that an LPPM achieving location privacy metrics such as geo-indistinguishability also satisfies $ \epsilon $-spatiotemporal event privacy.
We define $\epsilon$-spatiotemporal event privacy as the indistinguishability between an event and its negation because it provides a clear privacy semantics: the adversaries cannot infer or distinguish  whether the event happened or not. 
Alternatively we can define privacy as indistinguishability between an event and an alternative event.  
In addition to requiring users to define the alternative events, it presents additional challenges that may involve analyzing the correlations and exclusivity between events and complicated composition property of $ \epsilon $-spatiotemporal event privacy. 
We defer this to future work. }


\section{Quantifying $ \epsilon $-Spatiotemporal Event Privacy}
\label{sec: quantifying}

\subsection{Overview of our approach}
Given the emission matrix of an LPPM that takes input as user's true location $ u^t  $ and outputs a perturbed location $ o^t $ at time $ t $, quantifying the privacy guarantee under the metric of $ \epsilon $-spatiotemporal event privacy is equivalent to calculating the maximum ratio of $\frac{\Pr(o_1,o_2,\cdots,o_T|\textsc{Event})}{\Pr(o_1,o_2,\cdots,o_T|\lnot\textsc{Event})}$ given any $ o_1,o_2, \cdots,o_T $.
It is hard to derive this ratio directly from the emission matrix without specifying $ o_1,o_2,\cdots,o_T $ and correlations between locations.
Firstly, it is because we are not clear about the relationship between the perturbed output $ o^t $ and the defined \textsc{event}.
Secondly,  the predicates in the defined \textsc{event} may not be independent.
For example, if \textsc{event} is defined as {\footnotesize $ (u^1=s_1) \land (u^2=s_2)$}, \revA{the probability of such event, i.e.,} {\footnotesize$ \Pr((u^1=s_1) \land (u^2=s_2)) $}, may not be equal to {\footnotesize $ \Pr((u^1=s_1)) * \Pr((u^2=s_2)) $ } because the true locations in user's trajectory can be temporally correlated.
\revA{For simplicity, we first model the correlation between user's  consecutive locations using first-order\footnote{\revA{If the Markov model is high-ordered, i.e., the transition matrix has a larger state domain, our approach still works by applying the new matrix.}}  time-homogeneous\footnote{\revA{If the Markov model is  time-varying, i.e., transition matrices at different $ t $ are not identical, our approach still works. We explain this in the next section.}} Markov model, i.e., the transition matrix is identical at each $ t $.}
Then, we quantify this ratio w.r.t. given observations $ o_1,o_2, \cdots,o_T $ and a given user's initial probability $ \boldsymbol{\pi} $, so that we can directly calculate the $\Pr(o_1,o_2,\cdots,o_T|\textsc{Event}) $.
In Section \ref{sec:priste}, we will design a mechanism for  spatiotemporal event privacy w.r.t. any observations and arbitrary initial probability.
Our goal in this section is to calculate the likelihood of the observations given {\footnotesize $ \textsc{Event}  $} or {\footnotesize $ \lnot\textsc{Event}  $} , i.e., {\footnotesize $\Pr(o_1,o_2,\cdots,o_T| \textsc{Event}) $} or {\footnotesize $\Pr(o_1,o_2,\cdots,o_T| \lnot \textsc{Event}) $}, which can be derived by {\footnotesize $\Pr(o_1,o_2,\cdots,o_T|\textsc{Event}) = \frac{\Pr(o_1,o_2,\cdots,o_T, \textsc{Event}) }{\Pr( \textsc{Event}) } $}.
We call {\small $\Pr( \textsc{Event})  $} as \textit{prior probability} of the event, and {\small $\Pr(o_1,o_2,\cdots,o_T, \textsc{Event}) $} as \textit{joint probability} of the event.

One challenge of calculating the prior or joint probabilities of the event is the computational complexity.
Given an arbitrary spatiotemporal event, we need to enumerate all possible combination of the Boolean expression for prior and joint probabilities, which can be exponential to the number of predicates in the expression.
To address this problem, 
we propose a two-possible-world method for computing the prior and joint probabilities in Sections \ref{subsec:prior} and \ref{subsec:joint}, respectively.

For ease of exposition, we define notations frequently used in the following sections.
{\small $\textbf{M}\in\mathbb{R}^{m\times m}$} denotes a transition matrix that describes  temporal correlations in user's location.
At timestamp $1$, an initial probability is denoted by {\small $\boldsymbol{\pi}\in[0,1]^{1\times m}$}. During timestamp {\small $\{1, 2, \cdots, T\}$}, the probability of the true location {\small $\Pr(u^t)$} is denoted by a row vector {\small $\textbf{p}_t\in[0,1]^{1\times m}$} where the $i$th element denotes $\Pr(u^t=s_i)$. A Markov model follows the transition  property of {\small $\textbf{p}_{t+1}=\textbf{p}_t\textbf{M}$}, e.g., after a Markov transition, {\small $\textbf{p}_{2}=\boldsymbol{\pi}\textbf{M}$}  at timestamp $2$ given {\small $\textbf{p}_1=\boldsymbol{\pi}$}. 



The notations below for matrix computation are also used in the rest of this paper. 
Let $\textbf{0}$ and $\textbf{1}$ be  row vectors with $m$ elements being $0$ and $1$ respectively. 
$[\textbf{0},\textbf{1}]$ is a row vector in {\small $\mathbb{R}^{1\times 2m}$}. 
$\textbf{a}\circ\textbf{b}$ denotes the Hadamard product of $\textbf{a}$ and $\textbf{b}$. 
$\textbf{a}^\textbf{D}$ is a diagonal matrix with the elements of vector $\textbf{a}$ on the diagonal. 

\vspace{-5pt}
\subsection{Computing Prior Probability of an Event}
\label{subsec:prior}
To avoid the exponential complexity, we propose an efficient algorithm with two possible worlds.
The idea is to elaborate a ``new" transition matrix $ \textbf{M}_t \in \mathbb{R}^{2m\times 2m}$ at each time $ t $ which encodes the complex spatiotemporal event inside, so that the calculation of the prior or joint probability for a complicated event is the same as one simple predicate.

\noindent{\bf \textbf{Intuition.}} 
The main idea of our method is to use two virtual worlds denoting whether the \textsc{Event} is true or false.  
The states in the two worlds denote the joint probabilities $\Pr(u^t=s_i,\textsc{Event})$ and $\Pr(u^t=s_i,\lnot\textsc{Event})$. 
For \textsc{Presence}, once a trajectory enters into the region of the \textsc{Presence}, its probability will be kept in the world of true \textsc{Event} forever. 
For \textsc{Pattern}, the probability distribution among the two worlds are derived at the beginning timestamp of the \textsc{Event}, and only the trajectories satisfying the \textsc{Pattern} will be kept in the world of true \textsc{Event}. At last, the sum of probabilities in the world of true \textsc{Event} will be $\Pr(\textsc{Event}\textrm{ is true})$.

\begin{figure}[h]
	\centering
	\includegraphics[width=4cm]{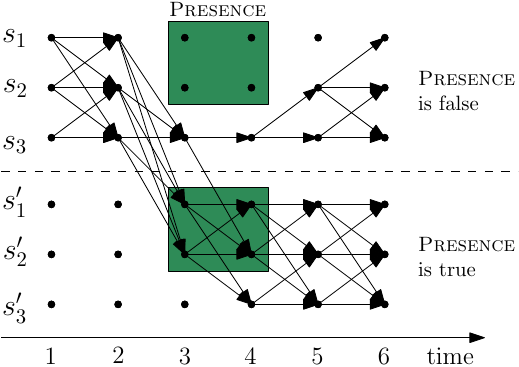}
	\caption{{\small New Markov transitions: all transitions going to the \textsc{Presence} area will be re-directed to the virtual worlds.}}
	\label{Figure-example-prior-presence}
\end{figure}

\noindent{\bf \textsc{Presence.}} We first study the $\textsc{Presence}$ event. Let us consider the following example. 
\begin{example}
	\label{example-prior-presence}
	Let $\mathbb{S}=\{s_{1},s_{2},s_{3}\}$.
	A \textsc{Presence} event is defined in $s_1$ and $s_2$ during  $t=3$ and $t=4$, i.e., $\textbf{s}=[1,1,0]^{\intercal}$, $start=3, end=4$.
	The transition matrix $\textbf{M}$ is given below. 
	\begin{myAlignSSS}
	\textbf{M}=\left[
	\begin{scriptsize}
	\begin{array}{ccc}
	0.1 &0.2&0.7\\
	0.4 & 0.1&0.5\\
	0&0.1&0.9\\
	\end{array}
	\end{scriptsize}
	\right]
	\end{myAlignSSS}
\vspace{-5pt}
	Then Fig.\ref{Figure-example-prior-presence}  shows the new transitions in the two worlds, the top world and the bottom world separated by the dashed line in Fig.\ref{Figure-example-prior-presence}. 
	From time $1$ to $2$, a normal transition can be made. At timestamp $2$, all the transitions going to the states $s_1$ and $s_2$ will be re-directed to the new states $s_1'$ and $s_2'$, denoting the states when the \textsc{Presence} happens. Other transitions that do not go the the area will perform normally. Similarly at time $3$, the transition from $s_3$ to $s_2$ will also go to the state $s_2'$ because the event is also true in this case. After time $4$, the original Markov transitions come back to work again. 
\end{example}

The intuition  can be formalized as follows. First, the original probabilities in $\mathbb{R}^{1\times m}$ is extended to $\mathbb{R}^{1\times 2m}$. Thus the initial probability $\boldsymbol\pi$ becomes $[\boldsymbol\pi,\textbf{0}]$.
Second, the transition matrix $\textbf{M}_{t}$ becomes the form of four transition matrices between the two virtual worlds, i.e. the \textsc{Event} is true or false, in Equation (\ref{new-M-0}). Then the new transition matrices can be derived in Equations (\ref{new-M-1}) and (\ref{new-M-2}) where $\textbf{M}$ is the original transition matrix and $\textbf{S}$ is the region of \textsc{Presence} defined in Definition \ref{def-presence}.
\begin{myAlignS}
\label{new-M-0}
\textbf{M}_t=
\left[
\begin{array}{cc}
\textrm{false}\rightarrow \textrm{false}&\textrm{false}\rightarrow \textrm{true}\\
\textrm{true}\rightarrow \textrm{false}&\textrm{true}\rightarrow \textrm{true}\\
\end{array}
\right] \textrm{on the event.}
\end{myAlignS}
\begin{myAlignS}
\label{new-M-1}
\textbf{M}_t=
\left[
\begin{array}{cc}
\textbf{M}-\textbf{M}\textbf{s}^{\textbf{D}}&\textbf{M}\textbf{s}^{\textbf{D}}\\
\textbf{0}^{\textbf{D}}&\textbf{M}\\
\end{array}
\right], start-1\leq t\leq end-1.
\end{myAlignS}
\begin{myAlignS}
\label{new-M-2}
\textbf{M}_t=
\left[
\begin{array}{cc}
\textbf{M}&\textbf{0}^{\textbf{D}}\\
\textbf{0}^{\textbf{D}}&\textbf{M}\\
\end{array}
\right], t<start-1\ or\  t\geq end.
\end{myAlignS}
Equation (\ref{new-M-1}), designed to capture and maintain all the transitions going to the region of the $\textsc{Presence}$, is the new transition matrix when entering (and inside) the event time. Equation (\ref{new-M-2}), designed to keep the original transitions in the two virtual worlds, is the new transition matrix when leaving (and before) the event time. Third, at the last time $T$, the probability of the $\textsc{Presence}$ will be the sum of all probabilities in the bottom world (where $\textsc{Presence}$ is true).

\begin{figure}[htbp]
	\centering
	\includegraphics[width=4cm]{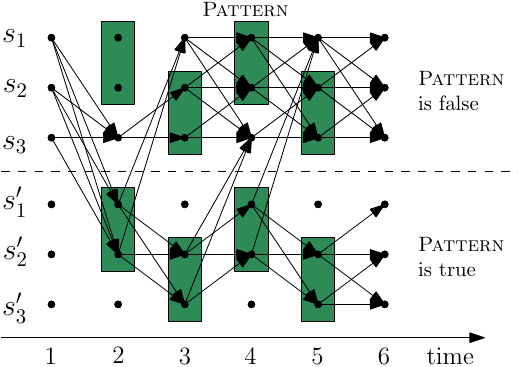}
	\caption{{\small New Markov transitions: at timestamp $1$, all transitions going to the defined area will be re-directed to the bottom world; at timestamp $2\sim 4$, only the transitions from the bottom world to the defined areas remain below.}}
	\label{Figure-example-prior-pattern}
\end{figure}

\noindent{\bf \textsc{Pattern.}} 
For \textsc{Pattern} events, the bottom world denoting the  event is true only needs to preserve the transitions going to the defined areas of the \textsc{Pattern} event. The following example shows the mechanism. 
\begin{example}
	\label{example-prior-pattern3}
	We study the \textsc{Pattern} event in Fig.\ref{Figure-example-prior-pattern}. 
	At time $1$, the transitions entering $s_1$ and $s_2$ go to $s_1'$ and $s_2'$. From time $2$ to $4$, the transitions in the above world perform normally. But the transitions from the bottom world go back to the top world if the destinations are not in  the defined regions. At time $5$, the  original Markov transitions come back to world again. 
\end{example}

From above example, the transition matrix for PATTERN only differs from that for PRESENCE  during the event time, shown in Equation (\ref{new-M-3}) where $\textbf{s}_{t}$ is the regions defined in Definition \ref{def-pattern}. 
The transitions outside the event time, i.e. {\footnotesize $t<start$} or {\footnotesize $t\geq end$},  will be the same as Equations (\ref{new-M-1}) and (\ref{new-M-2}). 
\begin{myAlignS}
\label{new-M-4}
\textbf{M}_t=
\left[
\begin{array}{cc}
\textbf{M}-\textbf{M}\textbf{s}^{\textbf{D}}&\textbf{M}\textbf{s}^{\textbf{D}}\\
\textbf{0}^{\textbf{D}}&\textbf{M}\\
\end{array}
\right], t=start-1.
\\
\label{new-M-3}
\textbf{M}_t=
\left[
\begin{array}{cc}
\textbf{M}&\textbf{0}^\textbf{D}\\
\textbf{M}-\textbf{M}\textbf{s}_t^{\textbf{D}}&\textbf{M}\textbf{s}_t^{\textbf{D}}\\
\end{array}
\right], start\leq t\leq end-1.
\\
\label{new-M-6}
\textbf{M}_t=
\left[
\begin{array}{cc}
\textbf{M}&\textbf{0}^{\textbf{D}}\\
\textbf{0}^{\textbf{D}}&\textbf{M}\\
\end{array}
\right], t<start-1\ or\  t\geq end.
\end{myAlignS}

In summary, the prior probability of any $\textsc{Event}$ can be derived as the sum of probabilities in the world where the $\textsc{Event}$ is true. Lemma \ref{theo-prior} shows the formal computation. 
\begin{lemma}
	\label{theo-prior}
	For initial probability $\boldsymbol\pi\in\mathbb{R}^{1\times m}$, the prior probability of an $\textsc{Event}$ of \textsc{Presence} and \textsc{Pattern}  is
	\begin{myAlignS}\Pr(\textsc{Event})=[\boldsymbol\pi , \textbf{0}] \prod_{i=1}^{end-1}\textbf{M}_i[\textbf{0},\textbf{1}]^\intercal\end{myAlignS}
	where $\textbf{M}_i$ is computed by Equations (\ref{new-M-1}), (\ref{new-M-2}), (\ref{new-M-4}), (\ref{new-M-3}), (\ref{new-M-6}).
\end{lemma}

\revA{When the transition matrices $ \textbf{M} $ at different $ t $ are not identical, the only extra effort is to re-compute Equations \eqref{new-M-1}$ \sim $\eqref{new-M-6} using the corresponding transition matrix $ \textbf{M} $ at $ t $.}

\vspace{-8pt}

\subsection{Computing Joint Probability of an Event}
\label{subsec:joint}
The calculation  of a spatiotemporal event and a set of observed locations, i.e, $ \Pr(o_1, o_2, \cdots, o _T, \textsc{EVENT}) $ is a little more complicated than previous sections since it dependents on not only the initial probabilities but also the emission matrix of the LPPM.
Similarly, we use two-possible-world method to avoid enumerating all possible cases of an event. 
We utilize  forward-backward algorithm\cite{schusterbockler_introduction_2007} to estimate the probability of the true state (true location) at timestamp $t$ given all observations $\Pr(u^t|o_1,o_2,\cdots,o_T)$. 
It first calculates a forward probability $\alpha_t^k=Pr(u^t=s_k,o_1,o_2,\cdots,o_t)$ iteratively, i.e., 
\begin{myAlignS}
\alpha_t^k=Pr(o_t| u^t=s_k)\sum_{i}\alpha_{t-1}^i Pr(u^t=s_k|u_{t-1}=s_i).
\end{myAlignS}
Then, a backward probability {\footnotesize $\beta_t^k=Pr(o_{t+1},o_{t+2},\cdots,o_T|u^t=s_k)$} can also be derived by
{ \begin{myAlignS}
{ \beta_t^k=\sum_i Pr(u_{t+1}=s_i|u^t=s_k)Pr(o_{t+1}|u_{t+1}=s_i)\beta_{t+1}^i}.
\end{myAlignS}}
By initializing $\beta_{T}^k=1$ for all $k$, we can obtain the estimation of $u^t$ as follows.
{ \begin{myAlignS}
{\footnotesize Pr(u^t=s_k|o_1,o_2,\cdots,o_T)=\frac{\alpha_t^k\beta_t^k}{\sum_i \alpha_t^i\beta_t^i}}
\end{myAlignS}}


\noindent{\bf Intuition.}
The intuition of our solution is to use the forward-backward algorithm in the two virtual worlds where the $\textsc{Event}$ is true and false. This is feasible because the emission probability, which determines the probabilities of the observations, is independent from any $\textsc{Events}$. Hence in our computation the forward probability and backward probability are $\Pr(\textsc{Event},o_1,o_2,\cdots,o_{t})$ for $t\leq end$ and $\Pr(o_{end+1},o_{end+2},\cdots,o_t|\textsc{Event})$ for $t>end$ respectively. 
By combining them together, we can obtain the posterior probability of the $\textsc{Event}$. Note that at any timestamp $t\leq end$, we do not see the future ($t>end$) observations. Thus the posterior probability only counts to the current timestamp $t$.

\noindent
\textbf{Before and During the Event.} 
In the forward algorithm, the probability 
$\alpha_t^k=\Pr(u^t=s_k,o_1,o_2,\cdots,o_t)$ is derived at timestamp $t$. We represent $\alpha_t^k$ in the vector form $\boldsymbol\alpha_t=[\alpha_1^1,\alpha_t^2,\cdots,\alpha_t^m]$. Then it can be derived as $\boldsymbol\alpha_t=(\boldsymbol\alpha_{t-1}\textbf{M}_{t-1})\circ\tilde{\textbf{p}}_{o_t}
=\boldsymbol\alpha_{t-1}\textbf{M}_{t-1}\tilde{\textbf{p}}_{o_t}^\textbf{D}$. Without any further observations, the joint probability can be derived from Lemma \ref{theo-prior}. The result is shown in Lemma \ref{lemma-post-before}.

\begin{lemma}
	\label{lemma-post-before}
	Given  initial probability $\boldsymbol\pi\in\mathbb{R}^{1\times m}$, the joint probability of an \textsc{Event} of $\textsc{Presence}$ or $\textsc{Pattern}$ and observations $o_1,o_2,\cdots,o_t$ at any timestamp $t\leq end$ is
	\begin{myAlignSSS}
	\label{eqn-post-before}
	\Pr(\textsc{Event},o_1,o_2,\cdots,o_t) 
	=[\boldsymbol\pi,\textbf{0}]\left(  \tilde{\textbf{p}}_{o_1}^\textbf{D}\prod_{i=2}^{t}(\textbf{M}_{i-1}\tilde{\textbf{p}}_{o_i}^\textbf{D})  \prod_{i=t}^{end-1}\textbf{M}_i   [\textbf{0},\textbf{1}]^\intercal \right)	\end{myAlignSSS}
\end{lemma}

\noindent
\textbf{After the Event.}
In the backward algorithm, {\footnotesize $\beta_t^k=\Pr(o_{t+1},o_{t+2},\cdots,o_T|u^t=s_k)$}. We represent it in the vector form {\footnotesize $\boldsymbol\beta_t=[\beta_t^1,\beta_t^2,\cdots,\beta_t^m]$}. 
Then it can be derived as {\footnotesize $\boldsymbol\beta_t=(\boldsymbol\beta_{t+1}\circ\tilde{\textbf{p}}_{o_{t+1}})\textbf{M}_t^\intercal 
=\boldsymbol\beta_{t+1}\tilde{\textbf{p}}_{o_{t+1}}^\textbf{D} \textbf{M}_t^\intercal$}  for any $t>end$. Similarly, the joint probability can be obtained by Lemma \ref{lemma-post-after}.
\begin{lemma}
	\label{lemma-post-after}
	Given  initial probability $\boldsymbol\pi\in\mathbb{R}^{1\times m}$, the joint probability of an \textsc{Event} of $\textsc{Presence}$ or $\textsc{Pattern}$ and observations $o_1,o_2,\cdots,o_t$ at any timestamp $t>end$ is
	\begin{myAlignSSS}
	\label{eqn-post-after}
	&\Pr(\textsc{Event},o_1,o_2,\cdots,o_t)=[\boldsymbol\pi,\textbf{0}] \nonumber \\ 
	&\left(  \tilde{\textbf{p}}_{o_1}^\textbf{D}\prod_{i=2}^{end}(\textbf{M}_{i-1}\tilde{\textbf{p}}_{o_i}^\textbf{D})\right) \left( [\textbf{1},\textbf{1} ]\prod_{i=t-1}^{end}(\tilde{\textbf{p}}_{o_{i+1}}^\textbf{D}\textbf{M}_i^\intercal)\circ [\textbf{0},\textbf{1}]\right)^{\intercal}
	\end{myAlignSSS}
\end{lemma}
\vspace{-10pt}


To summarize, now we can quantify the  ratio
 {\small $\Pr(o_1,o_2,\cdots,o_T|\textsc{Event}) = \frac{Pr(o_1,o_2,\cdots,o_T, \textsc{Event}) }{Pr( \textsc{Event}) } $} for spatiotemporal event privacy using Lemma \ref{theo-prior} to compute {\small $ \Pr( \textsc{Event}) $} and Lemmas \ref{lemma-post-before}, \ref{lemma-post-after} to compute {\small $ \Pr(o_1,o_2,\cdots,o_T, \textsc{Event}) $}.
 \revA{We note that our approach of computing the joint probability of an event is able to deal with different emission matrices at each $ t $. Since {\footnotesize $\tilde{\textbf{p}}_{o_t}$} is a vector of emission probabilities given the observation $o_t$, i.e, a column in the emission matrix, and {\footnotesize $\tilde{\textbf{p}}_{o_t}^\textbf{D}$} is a diagonal matrix whose diagonal elements are  {\footnotesize $\tilde{\textbf{p}}_{o_t}$},  we  only need to obtain {\footnotesize $\tilde{\textbf{p}}_{o_t}$}  and {\footnotesize $\tilde{\textbf{p}}_{o_t}^\textbf{D}$} from the corresponding emission matrix at $ t $, and then use such {\footnotesize $\tilde{\textbf{p}}_{o_t}^\textbf{D}$} in Equations \eqref{eqn-post-before} and \eqref{eqn-post-after}.}

\section{PriSTE framework}
\label{sec:priste}
In previous section, we designed methods for quantifying $ \epsilon $-spatiotemporal event privacy w.r.t. a specified  initial probability, which means that the quantification may not be true if attacker has a different initial probability since different initial probabilities may result in very different prior and joint probabilities.
In this section, we first solve this problem to ensure $ \epsilon $-spatiotemporal event privacy for any initial probabilities, then design a framework that converts planar Laplace mechanism into one  protecting spatiotemporal event privacy.

\subsection{Tackling Arbitrary Initial Probability}

According to the quantification in Section \ref{sec: quantifying}, we can calculate {\small $\frac{\Pr(o_1,o_2,\cdots,o_T|\textsc{Event})}{
		\Pr(o_1,o_2,\cdots,o_T|\lnot\textsc{Event})}$} given  $ o_1,o_2, \cdots,o_T $ and a given initial probability {$ {\boldsymbol\pi} $}.
In the next section, we will propose a framework that ensures any output from the framework satisfies $ \epsilon $-spatiotemporal event privacy. 
In this section, we show how to make sure the ratio is bounded given arbitrary initial probability.
Our idea to  is taking $ {\boldsymbol\pi} $ as a variable and solving the maximization problem of {\small $\frac{\Pr(o_1,o_2,\cdots,o_T|\textsc{Event})}{\Pr(o_1,o_2,\cdots,o_T|\lnot\textsc{Event})}- e^\epsilon $}.
We want to make sure the maximum value is always less than or equal to $0$, i.e., the user enjoys plausible deniability for her specified spatiotemporal event.

%
%

The following theorem shows the conditions related to $ {\boldsymbol\pi} $ that \revA{satisfies} $ \epsilon $-spatiotemporal event privacy.  
We will  formulate it as an optimization problem. 

\begin{theorem}
	\label{theo-eps-delta-DP}
	For an $\textsc{Event}$ of \textsc{Presence} or \textsc{Pattern} and an arbitrary initial probability $\boldsymbol\pi\in\mathbb{R}^{1\times m}$, 
	 $\epsilon$-spatiotemporal event privacy  is satisfied at any timestamp $t$, i.e., {\small $\frac{\Pr(o_1,o_2,\cdots,o_T|\textsc{Event})}{\Pr(o_1,o_2,\cdots,o_T|\lnot\textsc{Event})} \leq e^\epsilon$},
	if the observation $o_t$ is released based on the following two conditions
	\begin{myAlignSSS}
	\label{eqn-DP-check1}
	\boldsymbol\pi \left( [\textbf{1}^\textbf{D},\textbf{0}^{\textbf{D}}] \left( (e^\epsilon-1)\textbf{a}^{\intercal}\textbf{b}-e^{\epsilon}\textbf{a}^{\intercal}\textbf{c} \right)[\textbf{1}^\textbf{D},\textbf{0}^{\textbf{D}}]^{\intercal} \right) \boldsymbol\pi^\intercal 
	+\boldsymbol\pi[\textbf{1}^\textbf{D},\textbf{0}^{\textbf{D}}]{\left( \textbf{b}^\intercal \right)}\leq 0
	\end{myAlignSSS}
	\begin{myAlignSSS}
	\label{eqn-DP-check2}
	\boldsymbol\pi \left([\textbf{1}^\textbf{D},\textbf{0}^{\textbf{D}}] \left( (e^\epsilon-1)\textbf{a}^{\intercal}\textbf{b}+\textbf{a}^{\intercal}\textbf{c} \right)[\textbf{1}^\textbf{D},\textbf{0}^{\textbf{D}}]^{\intercal} \right)\boldsymbol\pi^\intercal
	-\boldsymbol\pi[\textbf{1}^\textbf{D},\textbf{0}^{\textbf{D}}]{\left( e^{\epsilon}\textbf{b}^\intercal\right)}\leq 0
	\end{myAlignSSS}
	 where
	\begin{myAlignSS}	\textbf{a}^\intercal=\prod_{i=1}^{end-1}\textbf{M}_i[\textbf{0},\textbf{1}]^\intercal
	\end{myAlignSS}
	For $t\leq end$,
	\begin{myAlignSS}
	\textbf{b}^\intercal= \tilde{\textbf{p}}_{o_1}^\textbf{D}\prod_{i=2}^{t}(\textbf{M}_{i-1}\tilde{\textbf{p}}_{o_i}^\textbf{D})  \prod_{i=t}^{end-1}\textbf{M}_i   [\textbf{0},\textbf{1}]^\intercal
	\textbf{c}^\intercal=  \tilde{\textbf{p}}_{o_1}^\textbf{D}\prod_{i=2}^{t}(\textbf{M}_{i-1}\tilde{\textbf{p}}_{o_i}^\textbf{D})    [\textbf{1},\textbf{1}]^\intercal 
	\end{myAlignSS}
	For $t>end$,
	\begin{myAlignSS}
	\textbf{b}^\intercal=\tilde{\textbf{p}}_{o_1}^\textbf{D}\prod_{i=2}^{end}(\textbf{M}_{i-1}\tilde{\textbf{p}}_{o_i}^\textbf{D}) \left([\textbf{1},\textbf{1} ]\prod_{i=t-1}^{end}(\tilde{\textbf{p}}_{o_{i+1}}^\textbf{D}\textbf{M}_i^\intercal) \circ [\textbf{0},\textbf{1}]\right)^\intercal
	\\
	\textbf{c}^\intercal=\tilde{\textbf{p}}_{o_1}^\textbf{D}\prod_{i=2}^{end}(\textbf{M}_{i-1}\tilde{\textbf{p}}_{o_i}^\textbf{D}) \left([\textbf{1},\textbf{1} ]\prod_{i=t-1}^{end}(\tilde{\textbf{p}}_{o_{i+1}}^\textbf{D}\textbf{M}_i^\intercal) \circ [\textbf{1},\textbf{1}]\right)^\intercal 
	\end{myAlignSS}
\end{theorem}

\vspace{-2pt}
\noindent{\bf Quadratic Programming.} 
To determine whether Equations (\ref{eqn-DP-check1}) and (\ref{eqn-DP-check2}) are true or not for arbitrary  $ \boldsymbol{\pi} $, we transform them to maximization problems: finding the maximum values of the left parts of Equations (\ref{eqn-DP-check1}) and (\ref{eqn-DP-check2}) under the constraints of $ 0 \leq p_i  \leq 1$ where $ p_i \in \boldsymbol{\pi}$.
As long as one of maximum values is larger than $ 0 $, then we know that the LPPM  (emission matrix) may not satisfy $ \epsilon $-spatiotemporal event privacy.
The maximization are equivalent to quadratic programing problem since they can be rewritten in a form of {\small $\boldsymbol\pi\textbf{A}\boldsymbol\pi^{\intercal}=\frac{1}{2}\boldsymbol\pi(\textbf{A}+\textbf{A}^{\intercal})\boldsymbol\pi^{\intercal}$} where  {\small \textbf{A} } is a matrix. 
 We skip  the computation details about solving such quadratic programing problem since many methods and tools  have been  proposed in  literature.
 In experiments, we use IBM CPLEX optimizer as our computation engine. 

\vspace{-5pt}
\subsection{PriSTE}
\vspace{-5pt}

\begin{figure}[h]
	\centering
	\includegraphics[width=6cm]{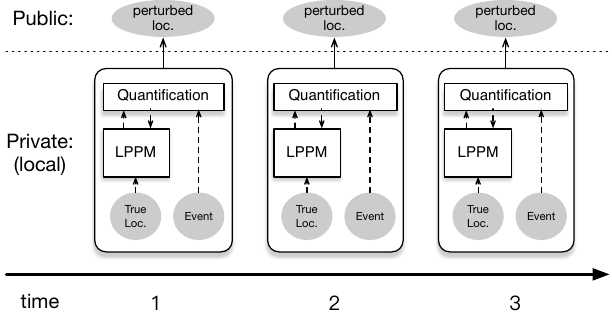}
	\caption{{\small PriSTE framework.}}
	\label{fig:fm}
\end{figure}

Based on the quantification techniques that we developed in previous sections, we propose a framework that converts a location privacy protection mechanism  into one protecting spatiotemporal event privacy.
The PriSTE framework is illustrated in Fig.\ref{fig:fm} and described in Algorithm \ref{alg-framework}.

The major components are \textit{Quantification} and a given LPPM.
Their interactions are described as follows.
First, the LPPM generates a perturbed location from the true location and pass it to the \textit{Quantification} component.
By Theorem \ref{theo-eps-delta-DP}, the \textit{Quantification} component check whether this perturbed location satisfies the ratio in Equation \eqref{eqn-eps-delta-DP1} ($ \epsilon $-spatiotemporal event privacy) or not,  under a sequence of previous observations and user-specified spatiotemporal events.
If not, we need to calibrate the emission matrix to ensure that it satisfies $ \epsilon $-spatiotemporal event privacy. 
The strategy of emission matrix calibration is  LPPM-dependent.
 In the next section, we demonstrate a case study for Planar Laplace Mechanism\cite{andres_geo-indistinguishability:_2013}, which is the state-of-the-art mechanism for location privacy.
 

\begin{algorithm}[h]
	\footnotesize
	\caption{PriSTE Framework}
	\begin{algorithmic}[1]
		\Require{ true location,
			$\epsilon$, LPPM, $\textbf{M}$, \textsc{Events}
		}
		\For{$t$ in $\{1,2,\cdots,T\}$}
		\State{generate $o_t$  with LPPM w.r.t. the true location;}
		\While{$\epsilon$-Spatiotemporal Event Privacy not hold}
		\label{alg-if-privacy}
		\State{\textit{calibrate} LPPM and generate $o_t$;}		
		\EndWhile
		\State{release $o_t$;}
		\EndFor
	\end{algorithmic}
	\label{alg-framework}
\end{algorithm}

\vspace{-10pt}
\revA{\subsection{Case Study 1: PriSTE with Geo-indistinguishability.}}
\vspace{-5pt}
\label{sec:case_study1}
In this section, we  instantiate PriSTE framework using $\alpha$-Planar Laplace mechanism ($\alpha$-PLM) which is designed for Geo-indistinguishability\cite{andres_geo-indistinguishability:_2013}.
We first show the computation details for quantifying $ \epsilon $-spatiotemporal event privacy by Theorem \ref{theo-eps-delta-DP}, and then design a greedy strategy for approximately achieving $ \epsilon $-spatiotemporal event privacy.
  
{\bf Algorithm Design.} 
To implement  the quantification component, we need to  (1) compute the internal parameters $\textbf{a}$, $\textbf{b}$ and $\textbf{c}$  shown in Theorem \ref{theo-eps-delta-DP}, and (2)  design a strategy to calibrate the emission matrix.

As for the calibration strategy for Planar Laplace Mechanism (PLM) with a specified privacy budget $ \alpha $ (which solely determine the shape of the output distribution), we  exponentially decay its privacy budget  because the smaller privacy budget implies stronger protection for location privacy and less information disclosure.
In our algorithm, decay rate $\frac{1}{2}$ for the privacy budget in line \ref{alg-budget-halve} of Algorithm \ref{alg-priv-check1} is a tunable parameter that provides a trade-off between efficiency and utility of the released locations.  
Setting a small value allows the algorithm converge faster, but at the cost of over-perturbing the location at each timestamp. In contrast, using a large value is less efficient but allows better utility to be achieved.

A natural question is that can we always find an $\alpha$ to release a perturbed location that satisfies Equation \eqref{eqn-eps-delta-DP1}. 
The answer is yes because $\alpha$ converges exponentially to $0$. When $\alpha=0$, it releases no useful information about the true location, i.e., uniformly returning a location without using user's true position.
It is easy to verify that the Equations \eqref{eqn-DP-check1} and \eqref{eqn-DP-check2} are always true in this situation. 

Algorithm \ref{alg-priv-check1} shows the computation process. 
To boost the efficiency of our algorithm, we use intermediate matrices $\textbf{A}$ and $\textbf{B}$ to facilitate the computation of $\textbf{b}$ and $\textbf{c}$. 
At time $1$, we initialize the variables as line $\ref{alg-line-t1-start}\sim \ref{alg-line-t1-end}$. 
At any time before and inside the $\textsc{Event}$, we compute the variables as line $\ref{alg-line-in-start} \sim \ref{alg-line-in-end}$. 
At any timestamps after the $\textsc{Event}$, the variables are derived as line $\ref{alg-line-out-start}\sim \ref{alg-line-out-end}$. 
Then we use quadratic programming methods to check Eq.(\ref{eqn-DP-check1}) and (\ref{eqn-DP-check2}) to decide whether to release the $o_{t}$ or not. 
If not, we generate a new $o_{t}$ with only half $\alpha$, and repeat the above process again.
Finally, we update the matrices $\textbf{A}$ and $\textbf{B}$ as line {\small $\ref{alg-line-AB-start}\sim\ref{alg-line-AB-end}$}. 
If $t=end$,  in line \ref{alg-line-in-start}, the product {\footnotesize $\prod_{i=t}^{end-1}\textbf{M}_i$} will be the identity matrix. 
In line \ref{alg-line-update-A}, {\footnotesize $\textbf{M}_{0}$} is the identity matrix when $t=1$.
We note that for multiple \textsc{Events}, Algorithm \ref{alg-priv-check1} can be executed multiple times for each \textsc{Event}. 
\begin{algorithm}[h]
	\footnotesize
	\caption{PriSTE with Geo-indistinguishability.}
	\begin{algorithmic}[1]
		\Require{
			$\epsilon$, \textsc{Event}, $\alpha$-PLM, $\textbf{M}_i$, $\forall i\in\{1,2,\cdots,T\}$
		}
		\For{$t$ in $\{1,2,\cdots,T\}$}
		\State{$o_t\gets$ $\alpha$-PLM;}\Comment{{\tt \scriptsize initial budget$=\alpha$}}
		\label{alg-line-ot}
		\If{$t==1$}
		\State{$\textbf{a}^\intercal\gets  \prod_{i=1}^{end-1}\textbf{M}_i[\textbf{0},\textbf{1}]^\intercal$}
		\label{alg-line-t1-start}
		\State{$\textbf{A}\gets\textbf{I} $}
		\Comment{{\tt \scriptsize identity matrix}}
		\State{$\textbf{B}\gets\textbf{I}$}
		\State{$\textbf{b}^\intercal\gets   \tilde{\textbf{p}}_{o_1}^\textbf{D} \textbf{a}^\intercal$}
		\State{$\textbf{c}^\intercal\gets \tilde{\textbf{p}}_{o_1}^{\intercal} $}
		\label{alg-line-t1-end}
		\ElsIf{$t<=end$}
		\Comment{{\tt \scriptsize before and during \textsc{Event}}}
		\State{$\textbf{b}^\intercal\gets \textbf{A}\textbf{M}_{t-1} \tilde{\textbf{p}}_{o_t}^\textbf{D} \prod_{i=t}^{end-1}\textbf{M}_i [\textbf{0},\textbf{1}]^\intercal$}
		\label{alg-line-in-start}
		\State{$\textbf{c}^\intercal\gets \textbf{A}\textbf{M}_{t-1}\tilde{\textbf{p}}_{o_t}^\textbf{D}  [\textbf{1},\textbf{1}]^\intercal$}
		\label{alg-line-in-end}
		\Else
		\Comment{{\tt \scriptsize after \textsc{Event}}}
		\State{$\textbf{b}^\intercal\gets \textbf{A}\left(([\textbf{1},\textbf{1} ]\tilde{\textbf{p}}_{o_t}^\textbf{D}\textbf{M}_{t-1}^\intercal \textbf{B})\circ[\textbf{0},\textbf{1}]\right)^\intercal$}
		\label{alg-line-out-start}
		\State{$\textbf{c}^\intercal\gets \textbf{A}\left(([\textbf{1},\textbf{1} ]\tilde{\textbf{p}}_{o_t}^\textbf{D}\textbf{M}_{t-1}^\intercal \textbf{B})\circ[\textbf{1},\textbf{1}]\right)^\intercal$}
		\label{alg-line-out-end}
		\EndIf
		\If{Equations (\ref{eqn-DP-check1}) and (\ref{eqn-DP-check2}) hold} \Comment{{\tt \scriptsize \revA{$\epsilon$ is used here.}}}
		\label{alg-line-if-check}
		\State{release $o_t$;}
		\Comment{{\tt \scriptsize okay to  release $o_t$}}
		\Else
		\State{$\alpha\gets \frac{\alpha}{2}$, goto Line \ref{alg-line-ot};}
		\label{alg-budget-halve}
		\Comment{{\tt \scriptsize halve the budget}}
		\EndIf
		\If{ $t\leq end$}
		\label{alg-line-AB-start}
		\State{$\textbf{A}\gets \textbf{A}\textbf{M}_{t-1}\tilde{\textbf{p}}_{o_t}^\textbf{D}$}
		\label{alg-line-update-A}
		\Comment{{\tt \scriptsize update $\textbf{A}$  by the real $o_{t}$}}
		\Else
		\State{$\textbf{B}\gets\tilde{\textbf{p}}_{o_t}^\textbf{D} \textbf{M}_{t-1}^\intercal  \textbf{B}$}
		\Comment{{\tt \scriptsize update $\textbf{B}$  by the real $o_{t}$}}
		\EndIf
		\label{alg-line-AB-end}
		\EndFor
		
		%
	\end{algorithmic}
	\label{alg-priv-check1}
\end{algorithm}

%
%

\noindent{\bf Complexity.} 
The internal parameters $\textbf{a}$, $\textbf{b}$ and $\textbf{c}$ in Algorithm \ref{alg-priv-check1} need $O(mT)$ time to be evaluated. 
The major computational cost lies in the quadratic program for checking Equations (\ref{eqn-DP-check1}) and (\ref{eqn-DP-check2}). 
The complexity will be determined by the quadratic matrix {\footnotesize $[\textbf{1}^\textbf{D},\textbf{0}^{\textbf{D}}]\textbf{a}^\intercal\textbf{c}[\textbf{1}^\textbf{D},\textbf{0}^{\textbf{D}}]^{\intercal}$}. 
If it is positive definite, then the complexity is $O(m^{3})$. 
Otherwise, with any negative eigenvalues, it will be NP-hard \cite{pardalos_quadratic_1991}. 
In our experiments, we use IBM CPLEX  which can provide globally optimal results for quadratic program but may need a long computation time.
We  use a \textit{conservative release} strategy to remedy this:
\revA{we use a threshold to limit the computation time of quadratic program for checking Eq.\eqref{eqn-DP-check1} and Eq.\eqref{eqn-DP-check2}.
It will not release a perturbed location unless we are sure that the equations are true.
Although it may lead to suboptimal solution in budget calibration, it always guarantees $\epsilon$-spatiotemporal event privacy since every released locations satisfy Eq.\eqref{eqn-DP-check1} and \eqref{eqn-DP-check2}.}

\noindent{\bf Privacy Analysis.}
PriSTE framework relies on a local model, i.e., the assumption that adversaries cannot obtain user's locally stored  information as shown in Fig.\ref{fig:fm}.
Although line \ref{alg-line-ot} may be executed more than once at a time point $ t $, Algorithm \ref{alg-priv-check1} still satisfies $ \alpha ' $-geo-indistinguishability where $ \alpha ' $ is the final privacy budget used for releasing $ o_t $ because that is the only observation of attacker at time $ t $.
If we remove the assumption of local model, the above statements may not be true since attacker may observe the internal states of the algorithm (which is the privacy goal of \textit{pan-privacy}\cite{dwork_pan-private_2010}).
Examples of  internal states includes  multiple  $ o_t $ tested at $ t $ or the final $ \alpha ' $ used in the algorithm.
Anothe assumption that may affect the privacy guarantee is the transition matrix \textbf{M}, which we use it to model the correlations between locations and assume that it is given.
It is  an interesting future work to quantify the change of privacy loss in terms of $ \epsilon $-spatiotemporal event privacy  if the ground truth of correlation is not the modeled one.
We defer this study to future work.
 
 \vspace{-10pt}

\revA{\subsection{Case Study 2: PriSTE with $\delta$-Location Set Privacy.}
\label{sec:case_study2}
To demonstrate the effectiveness of PriSTE framework, we also instantiate it using another  privacy metric \textit{$\delta$-location set privacy}\cite{xiao_protecting_2015}\cite{xiao_loclok:_2017}, which is proposed for obtaining better utility by taking advantage of temporal correlation between consecutive locations in user's trajectory.
The key idea is that hiding the true location in any impossible locations (e.g., whose probabilities are close to 0) is a lost cause because the adversary already knows the user cannot be there.
In other words, it restricts the output domain of the emission matrix to \textit{$\delta$-location set}, which is a set containing minimum number of locations that have prior probability sum no less than $1-\delta$.
A larger $\delta$ indicates weaker privacy guarantee.}

\revA{The privacy metrics of  $ \alpha $-geo-indistinguishability and $\delta$-location set privacy  are orthogonal because the former requires a specific ``shape'' of emission distribution and the latter restricts output domain of the emission distribution.
In \cite{xiao_protecting_2015}, Xiao and Xiong proposed a framework to achieve $\delta$-location set privacy using a given LPPM.
For ease of comparison, we use $ \alpha $-LPM as the underlying LPPM for $\delta$-location set privacy.}

\begin{algorithm}[h]
	\scriptsize
	\caption{\revA{PriSTE with $\delta$-Location Set Privacy.}}
	\revA{\begin{algorithmic}[1]
		\Require{
			$\epsilon$, \textsc{Event}, $\alpha$-PLM, $\textbf{M}_i$, $\forall i\in\{1,2,\cdots,T\}$, $\pi$, $\delta$, $ \mathbf{M}$.
		}
		\For{$t$ in $\{1,2,\cdots,T\}$}
		\State{$ \mathbf{p}_t^-  \gets \mathbf{p}_{t-1}^+ \mathbf{M} $;}\Comment{{\tt \scriptsize Markov transition}}
		\State{Construct $ \Delta \mathbf{X}_t $}\Comment{{\tt \scriptsize $\delta$-location set}}
		\State{$o_t\gets$ $\alpha$-PLM within $ \Delta \mathbf{X}_t $;}
		\State{the same as Lines 3 $ \sim $ 15 in Algorithm \ref{alg-priv-check1};}
		\If{Equations (\ref{eqn-DP-check1}) and (\ref{eqn-DP-check2}) hold} \Comment{{\tt \scriptsize $\epsilon$ is used here.}}
		\label{alg-line-if-check}
		\State{release $o_t$;}\Comment{{\tt \scriptsize okay to  release $o_t$}}
		\State{Derive posterior probability $ \mathbf{p}_t^+ $ by Eq.\eqref{eq:posterior};}
		\Else
		\State{$\alpha\gets \frac{\alpha}{2}$, goto Line 4;}
		\label{alg-budget-halve}
		\Comment{{\tt \scriptsize halve the budget}}
		\EndIf
		\State{the same as Lines 21 $ \sim $ 25 in Algorithm \ref{alg-priv-check1};}
		\EndFor
	\end{algorithmic}}
	\label{alg-priv-check2}
\end{algorithm}

\revA{In Line 2, when $ t=1 $, we have {\footnotesize $ \mathbf{p}_0^+ = \pi $}.
In Line 8, according to \cite{xiao_protecting_2015}, the posterior probability can be calculated by  Equation \eqref{eq:posterior} where {\footnotesize $ \mathbf{p}_t^+[j] $} and {\footnotesize $ \mathbf{p}_t^-[i] $} are the $ i $th elements in the corresponding probability vectors.
\begin{myAlignSSS}
	\mathbf{p}_t^+[i] = \Pr(u^t = s_i|o_t) = \frac{\Pr(o_t|u^t = s_i)*\mathbf{p}_t^-[i]}{\sum_{j} \Pr(o_t|u^t = s_j)*\mathbf{p}_t^-[j]} \label{eq:posterior}
\end{myAlignSSS}}
\vspace{-2pt}
\revA{The calculation of $\delta$-location set requires the initial probability $\pi$.
In experiments, we set $\pi$ to a uniform distribution.}

\vspace{-2pt}
\section{Experimental Evaluation}
In experiments, we verified that Algorithms \ref{alg-priv-check1}  and  \ref{alg-priv-check2}  can adaptively calibrate the privacy budget of Planar Laplace Mechanism (PLM) at each timestamp for $ \epsilon $-spatiotemporal event privacy.
We highlight the following empirical findings.
\begin{itemize}
\item A stricter LPPM can satisfy a certain level of spatiotemporal event privacy \textit{without} any change, whereas a more loose LPPM may need to reduce its privacy budget significantly for protecting the same event.
\item For achieving a specific $\epsilon$-spatiotemporal event privacy, a stricter LPPM is \textit{not} always better in terms of data utility.
\item \revA{If the user's transition matrix has a significant pattern, an LPPM may need a small privacy budge to achieve $ \epsilon $-spatiotemporal event privacy.}
\end{itemize}

\subsection{Experiment Settings and Metrics}
\noindent{\bf Dataset.} We used real-life and synthetic datasets in experiments.
	Geolife data \cite{zheng_geolife:_2010} was collected from $182$ users in a period of over three years.
	It recorded a wide range of users' outdoor movements, represented by a series of tuples containing latitude, longitude and timestamp.
	The user's entire trajectory is used to train the transition matrix $\textbf{M}$, e.g. with R package ``markovchain''. 
	
	We generated a synthetic trajectory and its transition probability matrix as follows. First, a map with $20*20$ cells is generated. Then, the transition probability from one cell to another is proportional to the two-dimensional Gaussian distribution with scale parameter $\sigma$. 
	\revA{Here, a smaller   $\sigma$ indicates that the user moves to the adjacent cells in a higher probability, i.e., the transition matrix has a more significant pattern.}
	Finally, we produced trajectories with $\revA{50}$ timestamps using such transition matrix to simulate  movement of a user.

%
%

\noindent{\bf Quadratic Programming.} We use the IBM CPLEX optimizer\footnote{https://www.ibm.com/analytics/data-science/prescriptive-analytics/cplex-optimizer} to find the globally optimal solution for the quadratic programming in Algorithm \ref{alg-priv-check1}. 
We adopt a  strategy of \textit{conservative release} as mentioned previously and limit  the computation time for each optimization to 1 seconds.

\noindent{\bf \textsc{Events}. }
We investigate  $\textsc{Presence}$ and $\textsc{Pattern}$ events, which are represented by two parameters $\textbf{S}$ and $\textbf{T}$. For example, {\footnotesize $\textsc{Presence}({\textbf{S}}=\{1:10\},\textbf{T}=\{4:8\})$} is \textsc{Presence} event denoting the user appears in the region of {\footnotesize $\{s_{1},s_{2},\cdots,s_{10}\}$} during timestamps {\footnotesize $\{4,5,6,7,8\}$}.

\noindent{\bf Utility Metrics.} 
We use two metrics to evaluate data utility.
\begin{itemize}
	\item
	Privacy budget $\alpha$ used in PLM, including  $\alpha$ at each timestamp (see Section \ref{sec:utility1}) and the average $\alpha$ during the whole time period (see Section \ref{sec:utility2}).
	The higher privacy budget indicates the higher utility.
	\item
	The Euclidean distance between the perturbed locations and the true locations.
	The smaller Euclidean distance indicates the higher utility. 
\end{itemize}
We run our algorithm  \revA{$100$} times and aggregate the results to calculate average privacy budget and  Euclidean distance.

\subsection{Utility at Each Timestamp}
\label{sec:utility1}
In this section, we show the utility (average privacy budget over \revA{100} runs) at each timestamp  for protecting {\footnotesize $\textsc{Presence}(\textbf{S}=\{1:10\},\textbf{T}=\{4:8\})$} and {\footnotesize $\textsc{Presence}(\textbf{S}=\{1:10\},\textbf{T}=\{16:20\})$}.

\begin{figure}[h]
	\centering
	\begin{subfigure}{0.24\textwidth}
		\centering
		\includegraphics[width=4.0cm]{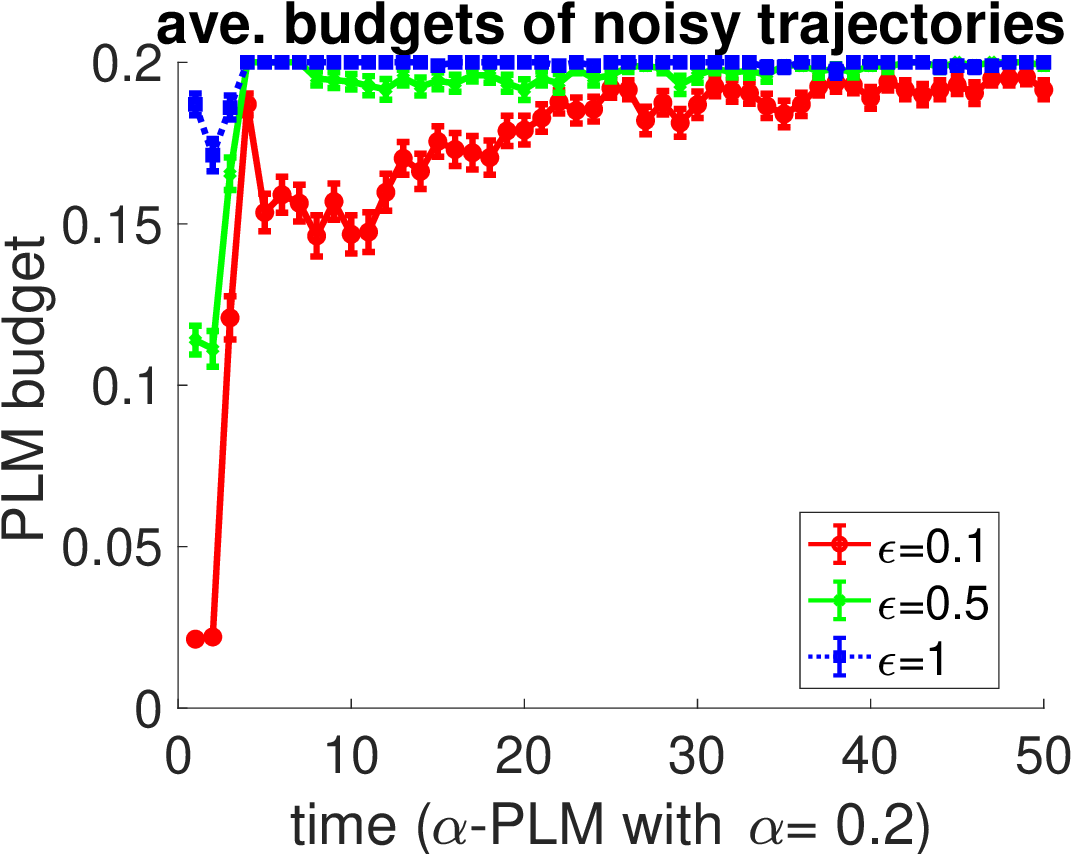}
		\vspace{-6pt}
		\caption{{\scriptsize $ 0.2 $-PLM  for different  $\epsilon$.}}
		\label{Figure-expmt-B11}
	\end{subfigure}
	\begin{subfigure}{0.24\textwidth}
		\centering
		\includegraphics[width=4.0cm]{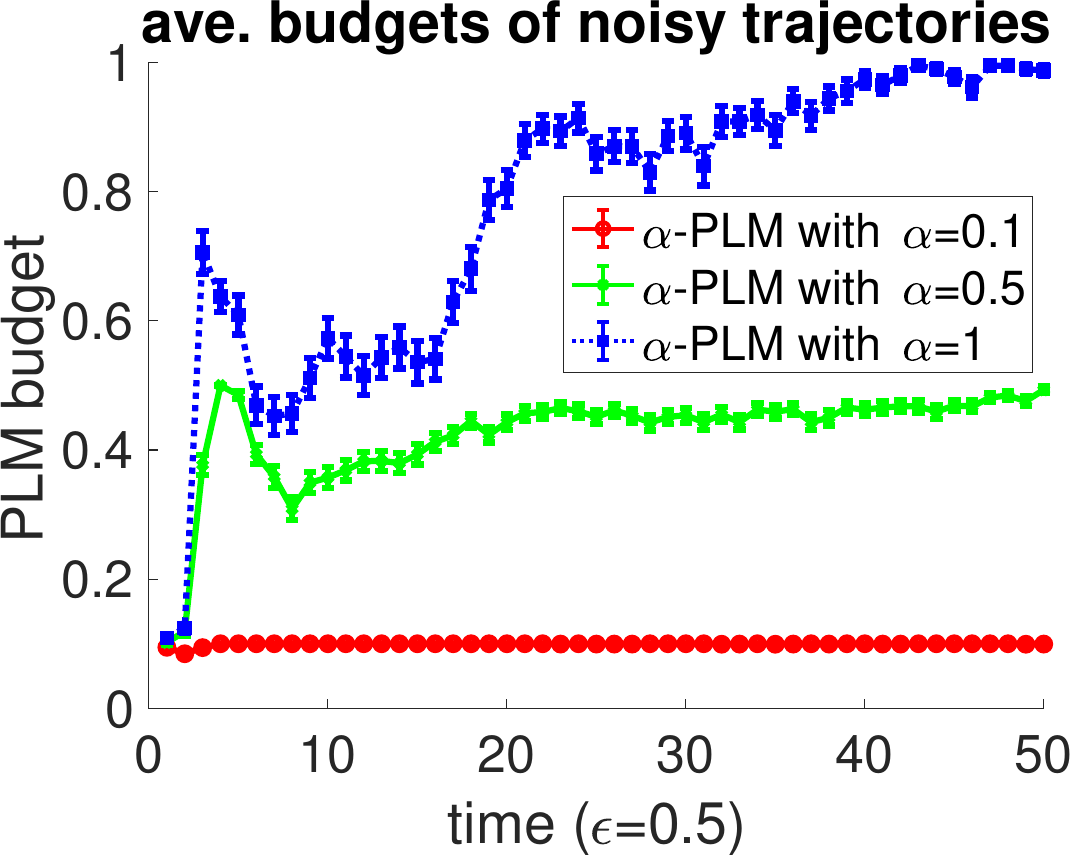}
			\vspace{-6pt}
		\caption{{\scriptsize  Different PLMs for $\epsilon=0.5$.}}
		\label{Figure-expmt-B22}
	\end{subfigure}
	\caption{{\revA{\footnotesize  $\textsc{Presence}(\textbf{S}=\{1:10\},\textbf{T}=\{4:8\})$ on synthetic data.}}}
	\label{Figure-expmt-B11-22}
\end{figure}

\begin{figure}[!htbp]
	\centering
	\begin{subfigure}{0.24\textwidth}
		\centering
		\includegraphics[width=4.0cm]{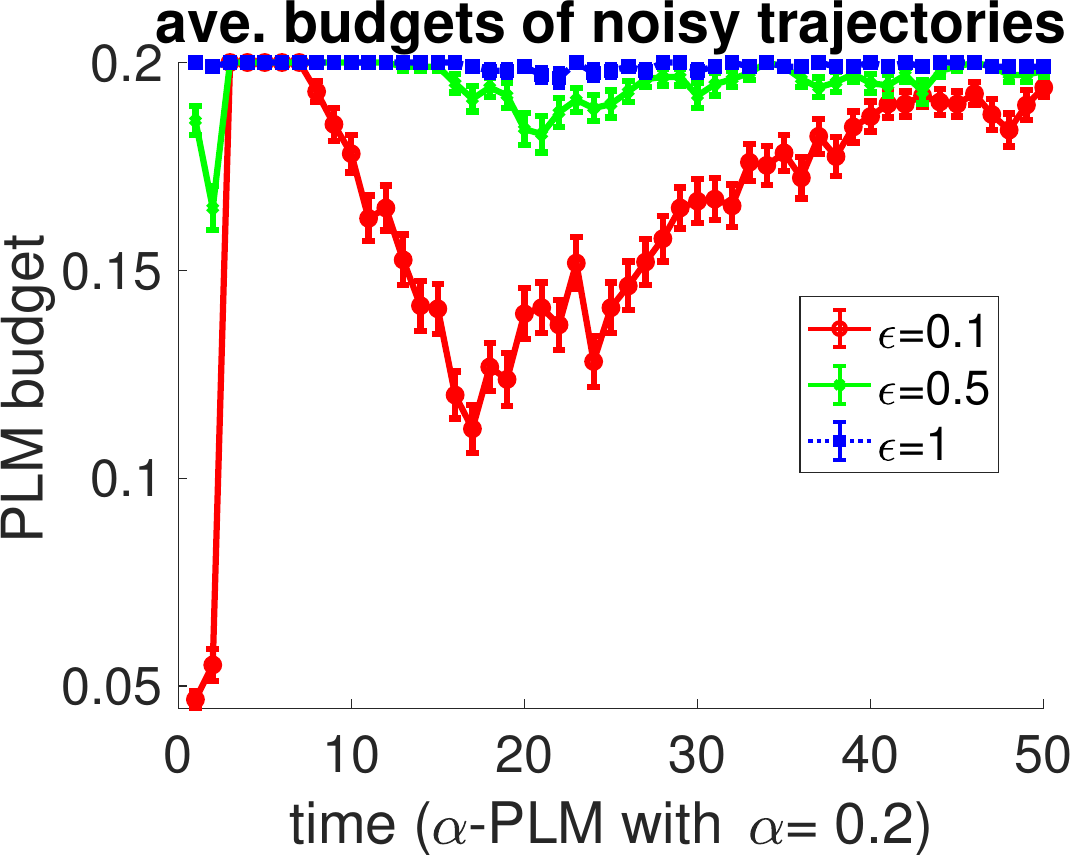}
			\vspace{-6pt}
		\caption{{\scriptsize $ 0.2 $-PLM  for different  $\epsilon$.}}
		\label{Figure-expmt-B33}
	\end{subfigure}
\begin{subfigure}{0.24\textwidth}
	\centering
	\includegraphics[width=4.0cm]{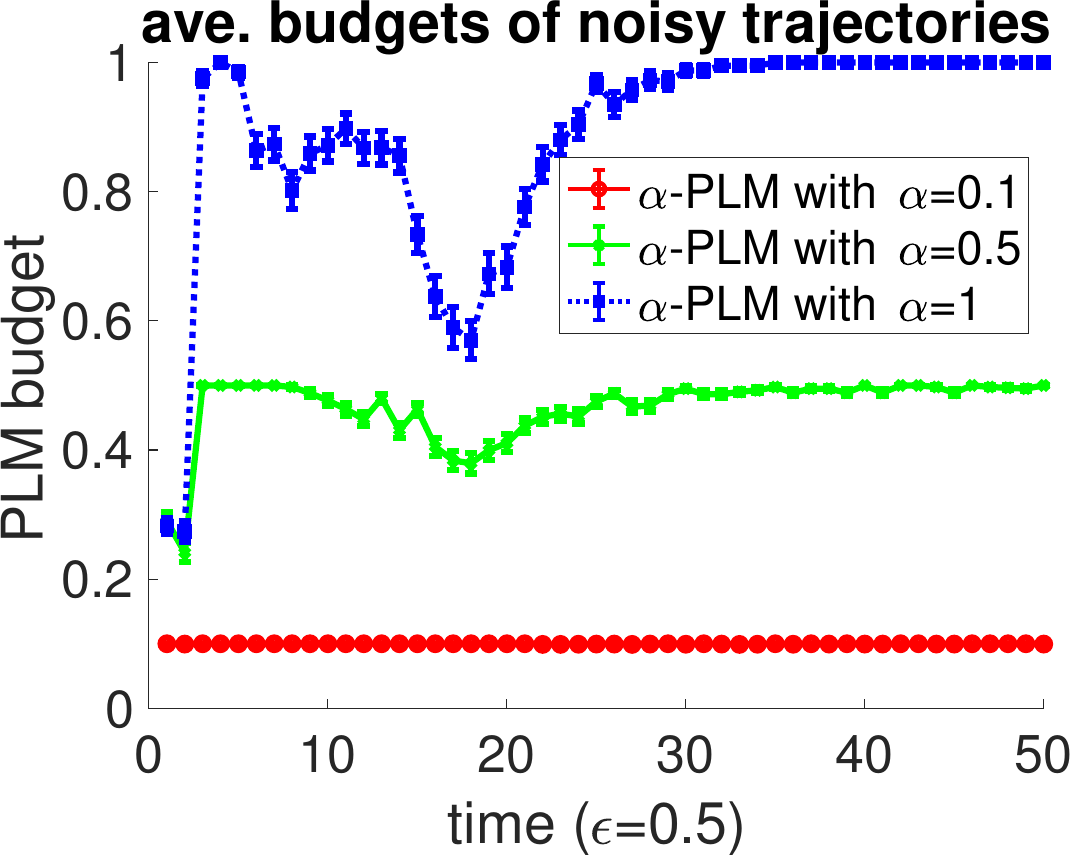}
		\vspace{-6pt}
\caption{{\scriptsize  Different PLMs for $\epsilon=0.5$.}}
	\label{Figure-expmt-B44}
\end{subfigure}
	\caption{\revA{{\footnotesize  $\textsc{Presence}(\textbf{S}=\{1:10\},\textbf{T}=\{16:20\})$ on synthetic data.}}}
	\label{Figure-expmt-B33-44}
\end{figure}

\revA{\textbf{PriSTE with Geo-indistinguishability.}}
 In Fig.\ref{Figure-expmt-B11-22}(a), it turns out that, 0.2-PLM  satisfies 1-spatiotemporal event privacy with only slight privacy budget reduction, and  satisfies 0.5-spatiotemporal event privacy with few budget reduction, but need to reduce more privacy budgets (to be stricter) in order to achieve 0.1-spatiotemporal event privacy.
 Similar results can be observed in Fig.\ref{Figure-expmt-B11-22}(b) and  Fig.\ref{Figure-expmt-B33-44}.
 \revA{We also observe that the standard deviation is larger for weaker LPPMs since these privacy budgets need to be frequently calibrated.}
 Hence, we can conclude that a stricter PLM for location privacy can protect spatiotemporal event without much calibration, but a more loose PLM may need to reduce its privacy budget significantly for $\epsilon$-spatiotemporal event privacy.

Comparing Fig.\ref{Figure-expmt-B11-22} with  Fig.\ref{Figure-expmt-B33-44}, where the events are defined on time periods 4$ \sim $8 and 16$ \sim $20 respectively,  we can see that privacy budgets  trend to be reduced during the defined time periods.
This indicates that the final $ \alpha $ used by PLM at each time point may disclose the definition of spatiotemporal event as we discussion in Section \ref{sec:case_study1}.
Hence, a local model is needed for PriSTE framework.

\begin{figure}[!htbp]
	\centering
	\begin{subfigure}{0.24\textwidth}
		\centering
		\includegraphics[width=4.0cm]{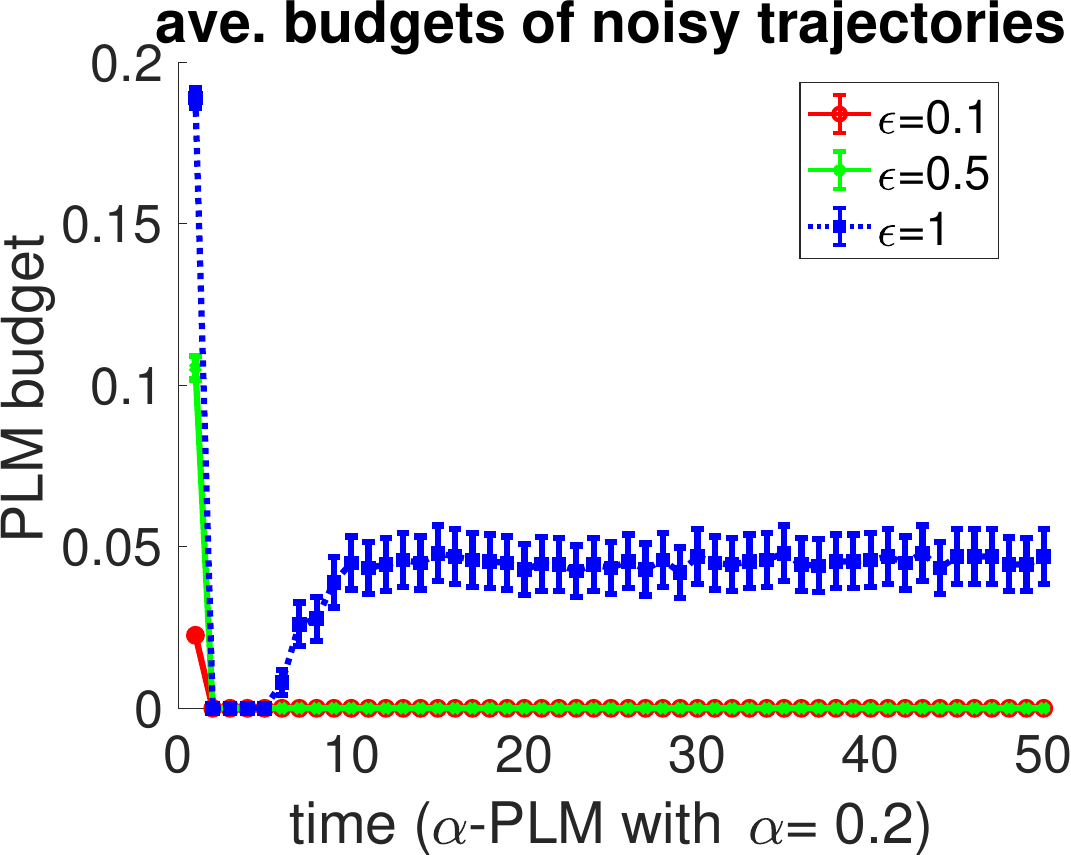}
		\caption{{\footnotesize $ 0.2 $-PLM  for different  $\epsilon$.}}
		\label{Figure-expmt-51}
	\end{subfigure}
	\begin{subfigure}{0.24\textwidth}
		\centering
		\includegraphics[width=4.0cm]{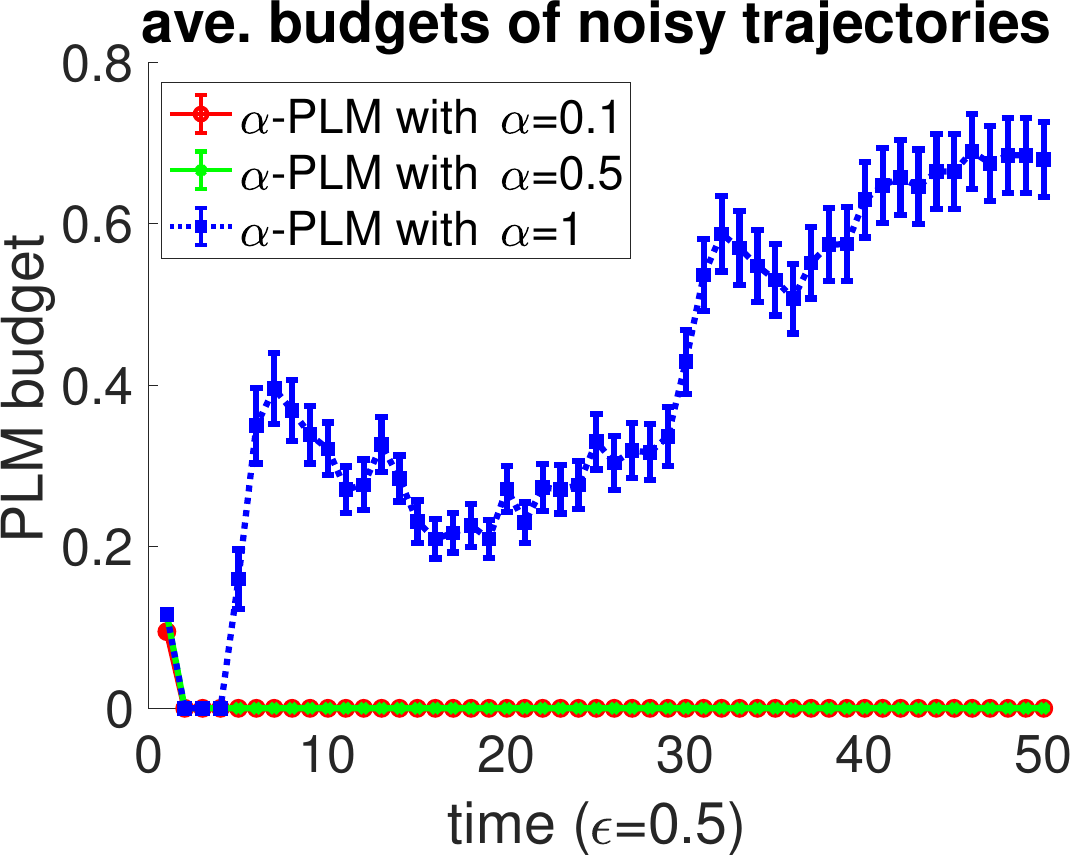}
		\caption{{\footnotesize Different PLMs for $\epsilon=0.5$.}}
		\label{Figure-expmt-B52}
	\end{subfigure}
	\caption{\revA{{\footnotesize  Protecting two events $\textsc{Presence}(\textbf{S}=\{1:10\},\textbf{T}=\{4:8\})$ and $\textsc{Presence}(\textbf{S}=\{1:10\},\textbf{T}=\{16:20\})$ on synthetic data.}}}
	\label{Figure-expmt-B5}
\end{figure}

%
\revA{\textbf{Protecting multiple events.}
Fig.\ref{Figure-expmt-B5} depicts  the  utilities when protecting two events sequentially using Algorithm \ref{alg-priv-check1}.
We can see that the utility is much worse than protecting each single event  in Fig.\ref{Figure-expmt-B11-22} or  Fig.\ref{Figure-expmt-B33-44}  because the algorithm needs to simultaneously check if $\epsilon$-spatiotemporal event privacy is satisfied for \textit{both} events at each time. 
If no perturbed location satisfying the privacy requirement of  both events simultaneously, the algorithm needs to halve the privacy budget  until finding an appropriate output.
}

\begin{figure}[h]
	\centering
	\begin{subfigure}{0.24\textwidth}
		\centering
		\includegraphics[width=4.0cm]{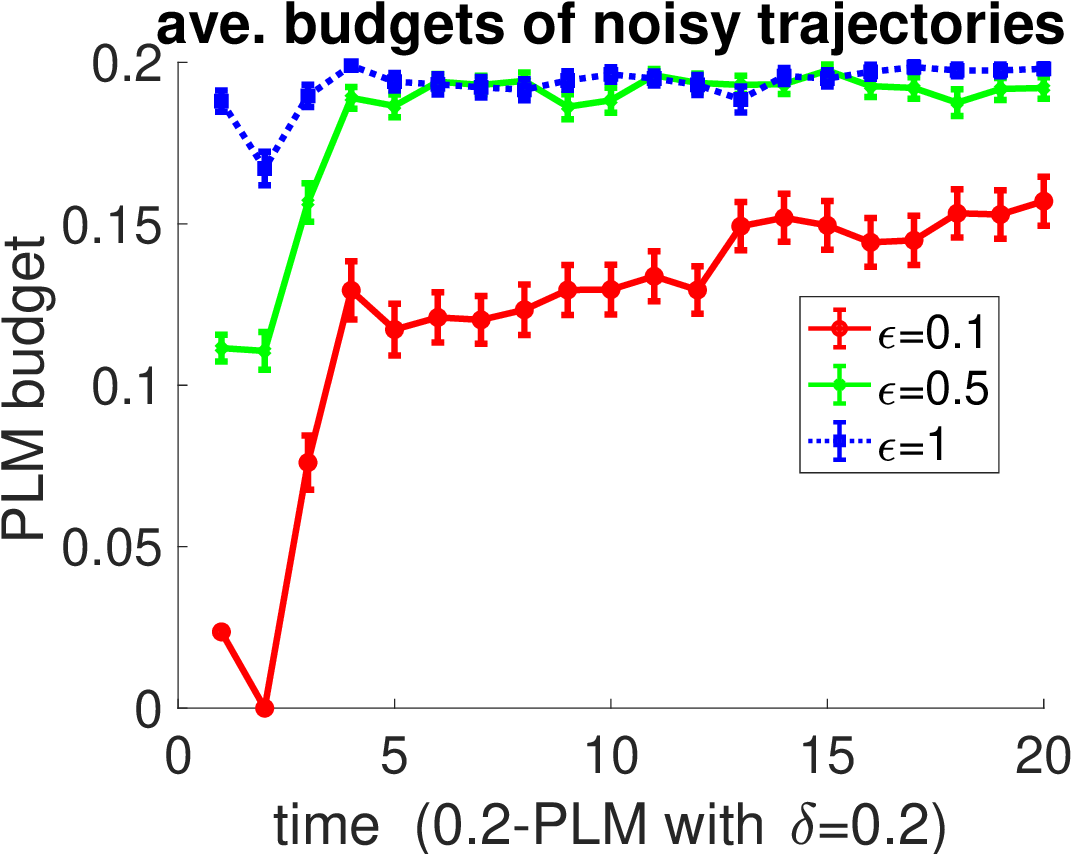}
		\vspace{-6pt}
		\caption{{\footnotesize $ 0.2 $-PLM with $ \delta $-location set privacy  for different  $\epsilon$.}}
		\label{Figure-expmt-61}
	\end{subfigure}
	\begin{subfigure}{0.24\textwidth}
		\centering
		\includegraphics[width=4.0cm]{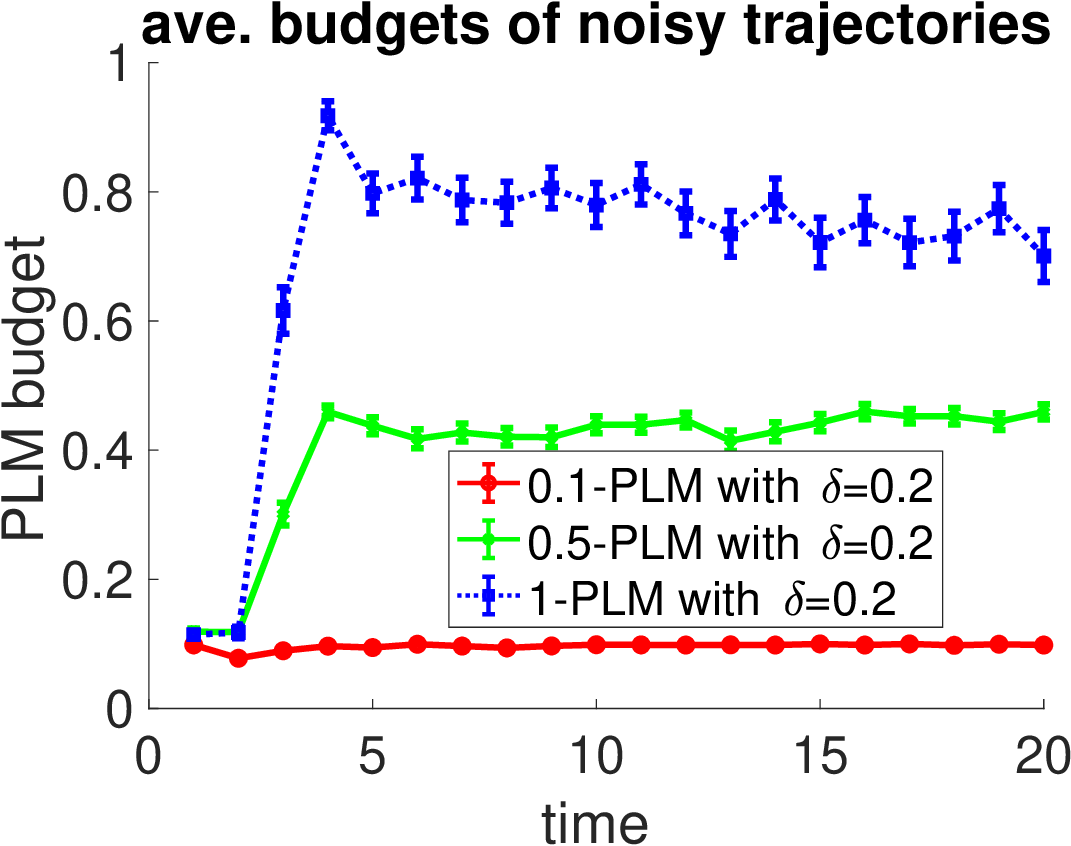}
		\vspace{-6pt}
		\caption{{\footnotesize  Different PLMs with $ \delta $-location set privacy for $\epsilon=0.5$.}}
		\label{Figure-expmt-B62}
	\end{subfigure}
	\caption{\revA{{\footnotesize  $\textsc{Presence}(\textbf{S}=\{1:10\},\textbf{T}=\{4:8\})$ on synthetic data.}}}
	\label{Figure-expmt-B6}
\end{figure}

\revA{\textbf{PriSTE with $\delta$-Location Set Privacy.}
In Fig.\ref{Figure-expmt-B6}, we show the utility of PriSTE with LPPMs that satisfy $\delta$-Location Set Privacy (Algorithm \ref{alg-priv-check2}).
Comparing  Fig.\ref{Figure-expmt-B6} with Fig.\ref{Figure-expmt-B11-22}, although both of them are using 0.2-PLM, the essential difference between them is the privacy metric: the former satisfies $\delta$-location set privacy and the latter satisfies geo-indistinguishability, i.e., 0.2-PLM in  Fig.\ref{Figure-expmt-B6} has a constrained output domain.
We can see that such  $0.2$-PLM in Fig.\ref{Figure-expmt-B6} has to reduce more privacy budgets to achieve $\epsilon$-spatiotemporal event privacy. 
Intuitively, it is because the privacy metric of $\delta$-location set privacy  implies a weaker privacy guarantee due to the temporal correlations and its LPPM has to be stricter (using a smaller privacy budget) for protecting the event.
} 

\vspace{-8pt}

\subsection{ Utility over Timestamps }
\label{sec:utility2}
\begin{figure}[h]
 	\centering
 	\begin{subfigure}{0.24\textwidth}
 		\centering
 		\includegraphics[width=4.0cm]{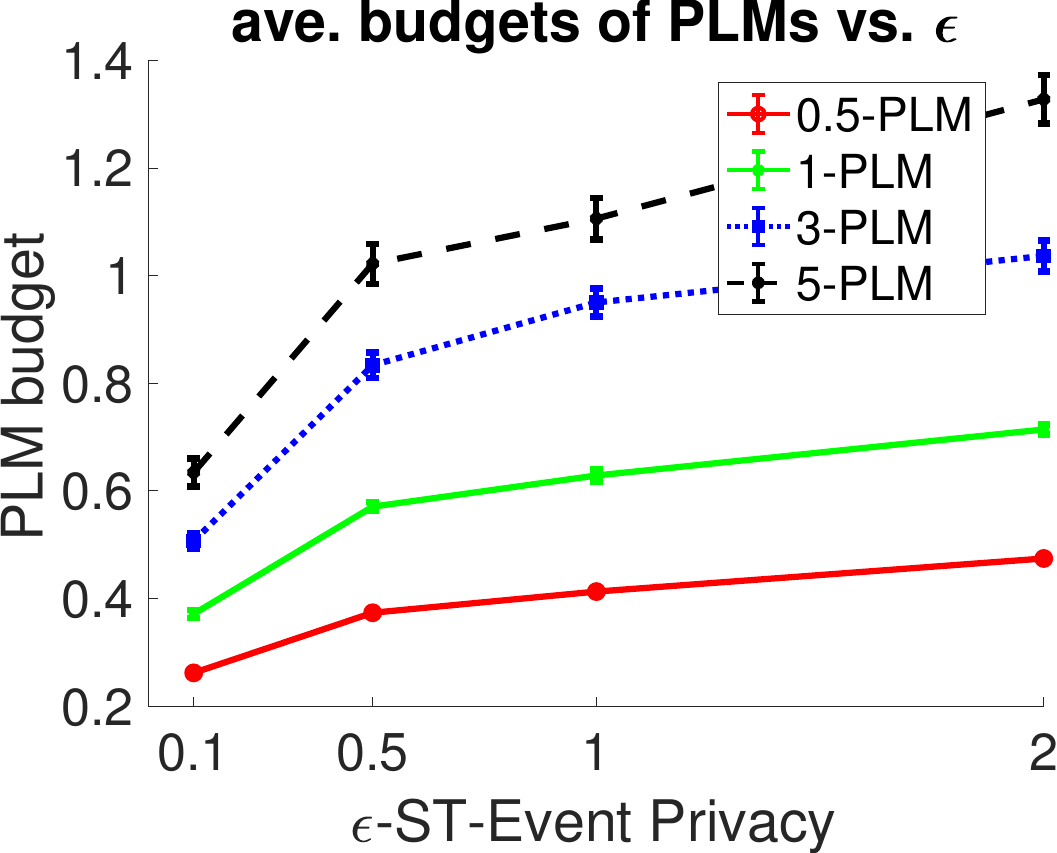}
 	\end{subfigure}
 	\begin{subfigure}{0.24\textwidth}
 		\centering
 		\includegraphics[width=4.0cm]{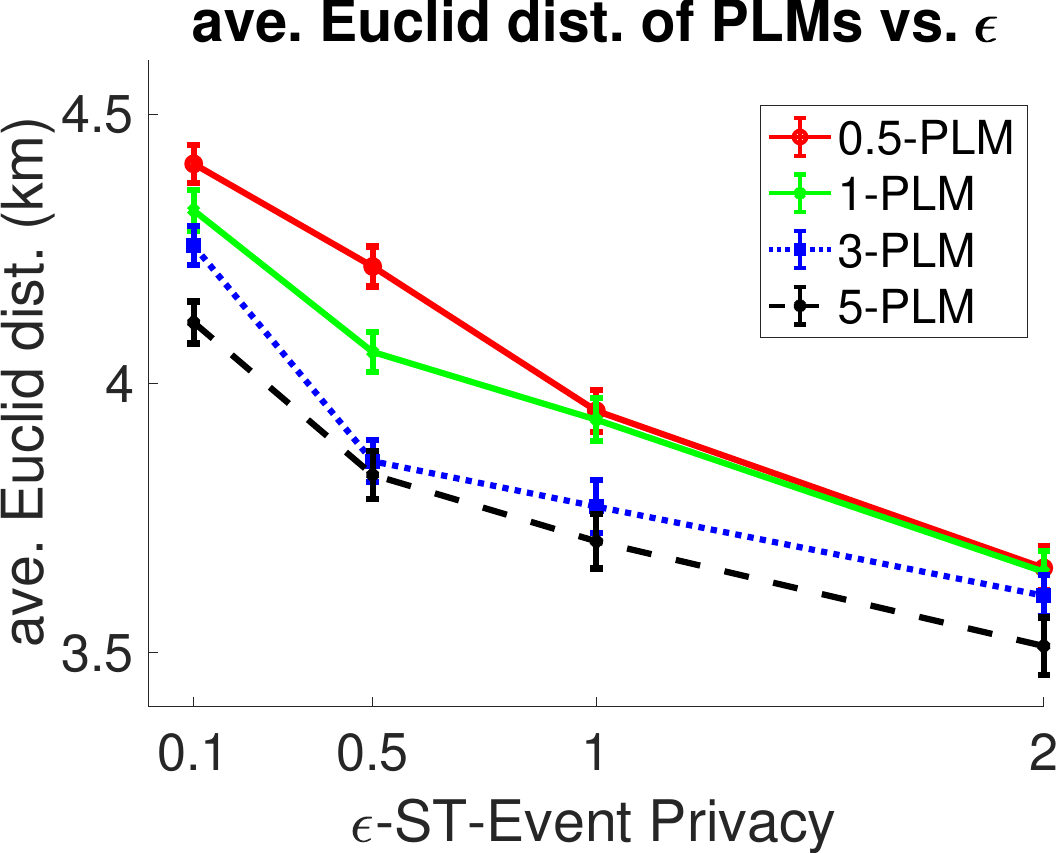}
 	\end{subfigure}
 	\caption{\revA{{\footnotesize  $\textsc{Presence}(\textbf{S}=\{1:10\},\textbf{T}=\{4:8\})$ on Geolife data.}}}
 	\label{Figure_C1_12}
 \end{figure}

\begin{figure}[!htbp]
	\centering
	\begin{subfigure}{0.24\textwidth}
		\centering
		\includegraphics[width=4cm]{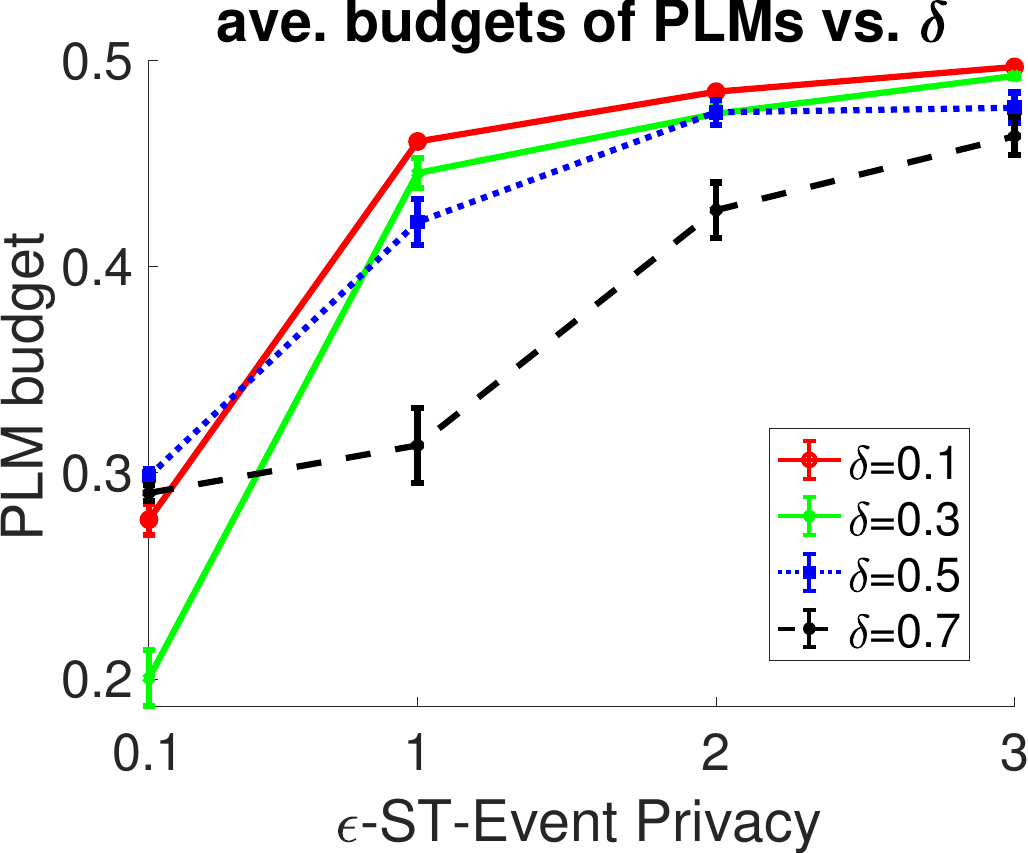}
		\label{Figure-expmt-C2-1}
	\end{subfigure}
	\begin{subfigure}{0.24\textwidth}
		\centering
		\includegraphics[width=4cm]{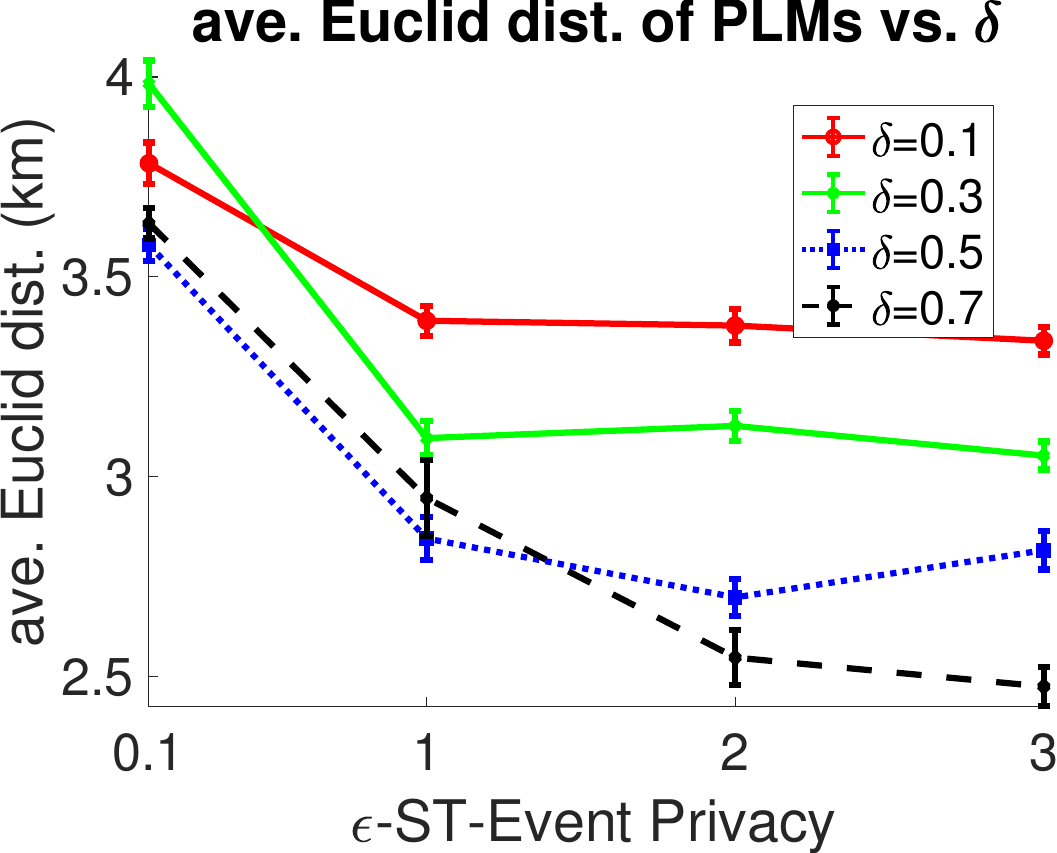}
		\label{Figure-expmt-C2-2}
	\end{subfigure}
	\caption{\revA{{\footnotesize  $\textsc{Presence}(\textbf{S}=\{1:10\},\textbf{T}=\{4:8\})$ on Geolife data (0.5-LPM with $ \delta $-location set privacy).}}}
	\label{Figure_C2_12}
\end{figure}

In this section,  we demonstrate the utility against different factors on the Geolife data and synthetic data.
Figures \ref{Figure_C1_12} and \ref{Figure_C2_12} are for protecting \textsc{Presence} event.
\revA{Due to space limitation, the  results of protecting  \textsc{Pattern} event are included in Appendices.}
Different from the utility in previous section which is averaged at each time, this section displays  the utility  that is further averaged over time points.
Hence, in the left parts of Figures \ref{Figure_C1_12} and \ref{Figure_C2_12} (ave. budget), the steeper lines indicate the budget may be reduced heavily at some timestamps.
 Generally, the utility increases with a larger $\epsilon$ in Fig.\ref{Figure_C1_12} and Fig.\ref{Figure_C2_12}.
 
 \revA{\textbf{Utility vs. $\alpha$-geo-indistinguishability.}
 In Fig.\ref{Figure_C1_12}, we can see that a larger $\alpha$-PLM needs to be calibrated heavily (i.e., steeper) for a small $\epsilon$.
 Interestingly,   PLMs who have larger average budgets (in the left figures) may not necessarily have better utility in terms of Euclidean distance.
 For example, we can see that,  at $ \epsilon=0.5$, the  Euclidean distance of  5-PLM and 3-PLM are very close; at $\epsilon=1 $ or $ 2 $,   0.5-PLM and 1-PLM appear to have almost  the same Euclidean distance.
 The reason is that PLMs who have  larger \textit{average}  budgets may have extremely small budgets at some timestamps, which results in the higher  \textit{average} Euclidean distance  averaged over time points.}


 \revA{\textbf{Utility vs. $\delta$-location set privacy.}
In Fig.\ref{Figure_C2_12}, we can see that 	 a PLM with  a larger $\delta$ tends to have a smaller average budget.
It is because the PLM with a larger $\delta$ indicates  a weaker privacy metric.
Hence, the PLM needs to be stricter (i.e, a small budget) to achieve spatiotemporal event privacy.
However, such PLM may have a better utility in terms of Euclidean distance as shown in right figure of Fig.\ref{Figure_C2_12}.
The reason is that  $\delta$-location set privacy with a larger $\delta$ restricts the output domain significantly, which makes perturbed location close to the true location in a high probability.
The results are in line with  the main purpose of $\delta$-location set privacy:  to have a better trade off between utility and privacy.
}

%
%

\begin{figure}[h]
	\centering
	\begin{subfigure}{0.24\textwidth}
		\centering
		\includegraphics[width=4cm]{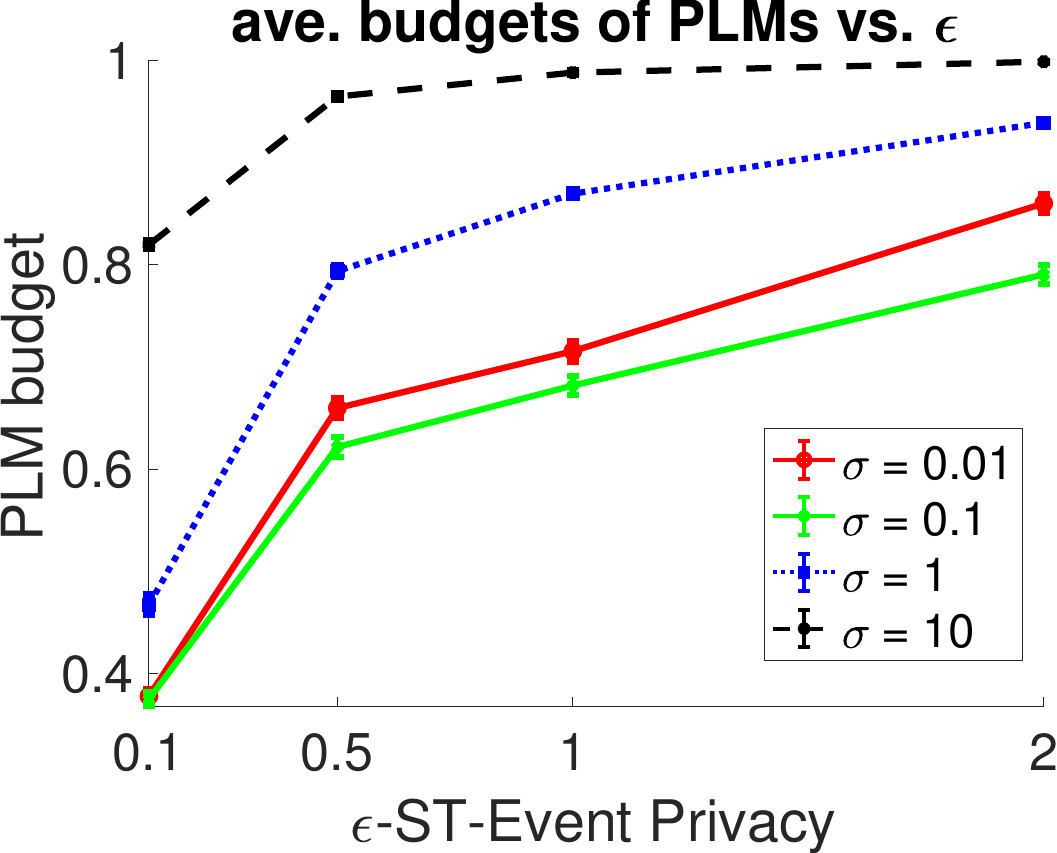}
		\label{Figure_C4_2}
	\end{subfigure}
	\begin{subfigure}{0.24\textwidth}
		\centering
		\includegraphics[width=4cm]{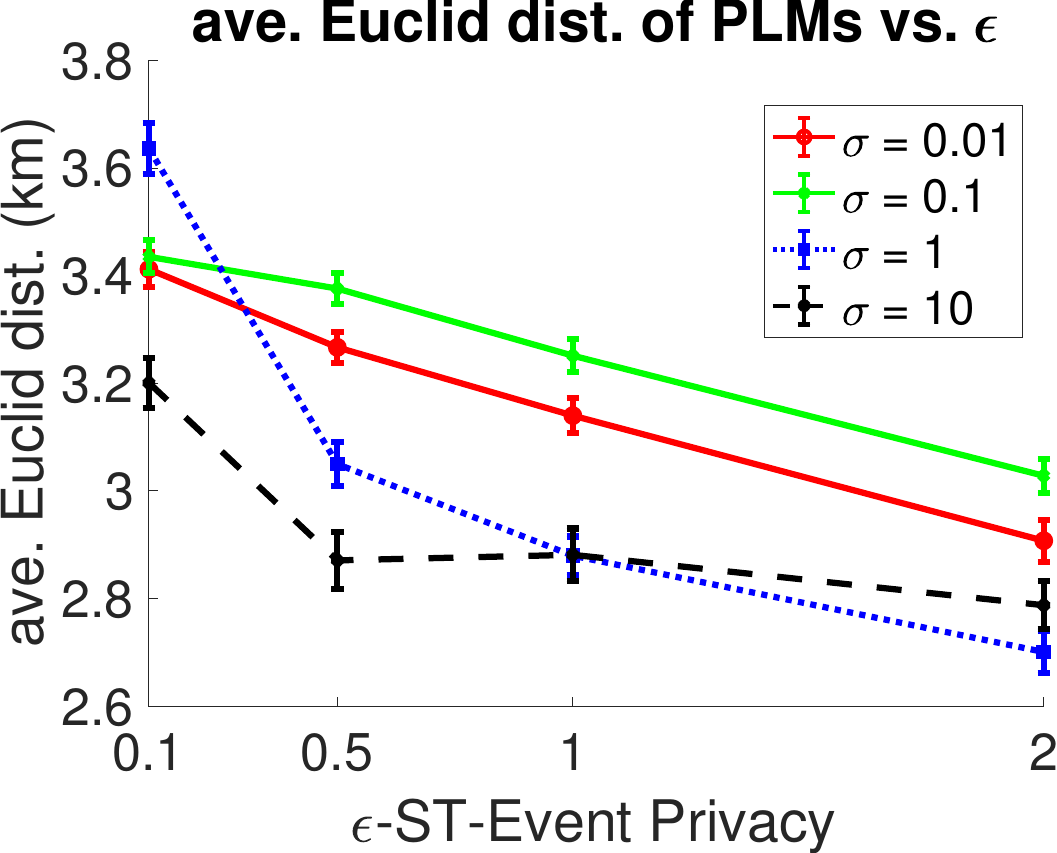}
		\label{Figure_C4_1}
	\end{subfigure}
	\caption{\revA{{\footnotesize  $\textsc{Presence}(\textbf{S}=\{1:10\},\textbf{T}=\{4:8\})$ on synthetic data (1-LPM with geo-indistinguishability).}}}
\label{Figure_C4_12}
\end{figure}

 \revA{\textbf{Utility vs. Transition Matrices.}
 We compare the utility against  transition matrices that have different strength of mobility patterns.
As we explained previously, a smaller $\sigma$ indicates a more significant mobility pattern.
Fig.\ref{Figure_C4_12} shows that, for the same LPPM, it is hard to protect a spatiotemporal event if user's mobility pattern is significant, i.e., the LPPM needs to be very strict by using a small privacy budget.
We also observe that there is no best LPPM for all $\epsilon$-spatiotemporal event privacy in terms of Euclidean distance.}
 	
\vspace{-10pt}
\subsection{Runtime}
\label{sec:rt}
\revA{
We name the size of $ \mathbf{T} $ and the size of $ \mathbf{S} $ as  \textit{event length} and \textit{event width}, respectively.
We also report the performance evaluation on \textit{conservative release} described  in Section \ref{sec:case_study1}.}

\begin{figure}[h]
	\centering
	\includegraphics[width=8.5cm]{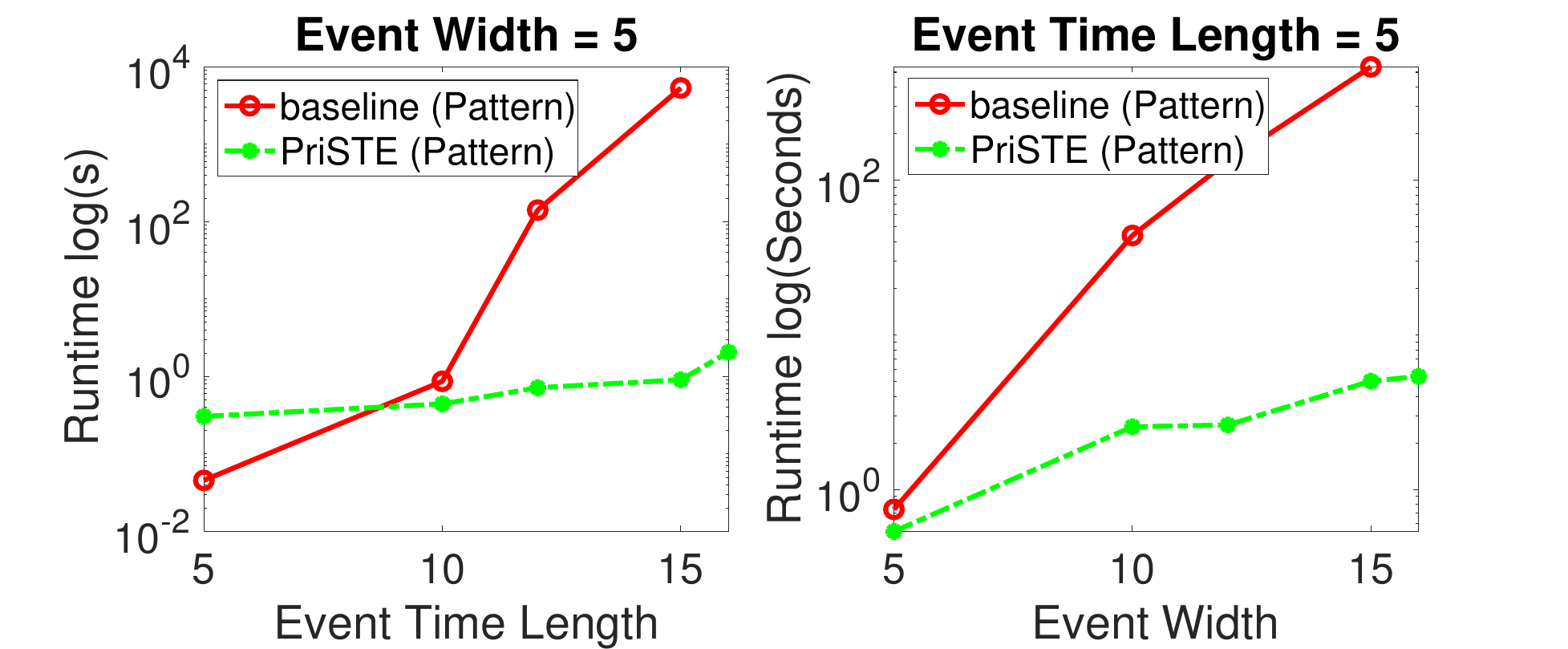}
	\caption{{\small Runtime}}
	\label{Figure-expmt-runtime}
\end{figure}

\noindent\revA{\textbf{Runtime vs. Event Length}.
We fix the event width as 5 and test 100 random events with length ranging from 5 to 15.
Fig.\ref{Figure-expmt-runtime} shows that the average runtime of the baseline  is exponential to event length and the runtime of our method  is linear to  the event length.}

\noindent\revA{\textbf{Runtime vs. Event Width}.
We fix the event length as 5 and test 100 random events with width ranging from 5 to 15.
Fig.\ref{Figure-expmt-runtime} shows that the average runtime of the baseline  is exponential to event width, while our method is polynomial to the event width, which is in line with the  complexity of {\small $ O(m^3)$}.}

\noindent\revA{\textbf{Runtime vs. Conservative Release}.
In Line 16 of Algorithm \ref{alg-priv-check1}, we set a threshold runtime in solving the quadratic program.
We do not release the perturbed location unless we are sure that Eq.\eqref{eqn-DP-check1} and Eq.\eqref{eqn-DP-check2} are true.
The threshold is a trade-off between runtime and utility as shown in Table \ref{tbl_runtime} among 100 runs.
We note that each runtime in Table  \ref{tbl_runtime}  includes the whole process of Algorithm \ref{alg-priv-check1}.
In our implementation, we set the threshold to 1 second.
We can see as the threshold increases, the number of conservative releases decreases, which results in increasing runtime.
On the other hand, the calibrated privacy budgets increasse as the threshold increases. 
This verifies the tradeoff between runtime and utility that can be achieved by the conservative release.}

\begin{table}[h]
	\centering
	\scriptsize
\revA{	\begin{tabular}{|p{35pt}|p{35pt}|p{35pt}|p{35pt}|p{35pt}|}
		\hline
			\textbf{threshold (s)} & \textbf{ave. total runtime (s)} & $\# $ \textbf{of}  \space\space \textbf{Conservative} \textbf{Release} &  \textbf{ave.}  \textbf{privacy} 
			\textbf{budget}& \textbf{ave.}  \textbf{Euclidean} \textbf{dist. (km)}
		 \\\hline
		0.01 & 1.1 & \textbf{33} &  0.16 & 2.22 \\\hline
		0.1 & 2.6 & 30 &  0.23 & 1.51 \\\hline
			1 & 5.9 & 21 &  0.22 & 1.52 \\\hline
		2 & 10.4 & 12 &  {0.29} & \textbf{0.93 }\\\hline
		5 & {19.5} & 8 &  0.27 & 1.41 \\\hline
		none  & \textbf{52.5} & 0 &  \textbf{0.31} & 0.97 \\\hline
	\end{tabular}}
	\caption{Runtime vs. threshold}
	\label{tbl_runtime}
\end{table}

\vspace{-10pt}

\section{Related Works}

\subsection{Location Privacy Preserving Mechanisms}

The LPPMs \cite{andres_geo-indistinguishability:_2013}\cite{xiao_protecting_2015}\cite{ghinita_preventing_2009}\cite{ardagna_obfuscation-based_2011}\cite{hwang_novel_2012} generally use some
obfuscation methods, like spatial cloaking, cell merging, location precision reduction or dummy cells, to manipulate the probability distribution of
users' locations. 
As differential privacy becomes a standard for privacy protection, \cite{andres_geo-indistinguishability:_2013} proposed a Geo-indistinguishability notion based on differential privacy and a planar Laplace mechanism to achieve it. Xiao et al.  \cite{xiao_protecting_2015}\cite{xiao_loclok:_2017} studied how to protect location privacy under temporal correlations with an optimal differentially private mechanism. 
\revA{Rodriguez-Carrion et al. \cite{rodriguez-carrion_entropy-based_2015}  also studied the effect of temporal dependencies on entropy-based location privacy metric.
They proposed a new privacy metric \textit{entropy rate} and perturbative mechanisms based on it, which can be an alternative LPPM in our framework for protecting spatiotemporal event privacy.}
Several studies \cite{shokri_quantifying_2011} \cite{bordenabe_optimal_2014} \cite{shokri_privacy_2015}  tried to achieve an optimal trade-off between the utility of applications and the privacy guarantee in the LPPMs.
Overall, above works all focused on the mechanisms of location privacy, which can be used in our framework as given LPPMs. Whereas we study a new problem of spatiotemporal event privacy. 

\subsection{Inferences on Location}
Various inference attacks can be carried out based on location information and external information such as moving patterns. In the aggregated setting, recent works have studied location or trajectory recovery attacks from aggregated location data\cite{liu_location_2018}  \cite{xu_trajectory_2017}
 or proximity query results from location data \cite{argyros_evaluating_2017}.
We mainly discuss the individual setting that is closely related to our work.
Studies
\cite{shokri_quantifying_2011}\cite{cao_quantifying_2017} investigated the question of how to formally quantify the privacy of existing LPPMs, given an adversary who can model users' mobility using a Markov process learned from population.  
\cite{liao_learning_2004} used a hierarchical Markov model to learn and infer a user's trajectory based on the places and temporal patterns they visited.  \cite{qiao_putmode:_2010} used the Continuous Time Bayesian Networks to predict uncertain trajectories of moving objects.
\revA{Li et al. \cite{li_swarm:_2010} uses frequent mining approach to find  moving objects that move within arbitrary shape of clusters for certain timestamps that are possibly nonconsecutive.}

Our privacy check algorithms for computing the prior and posterior probabilities of a spatiotemporal event given perturbed locations are inspired by the work in \cite{emrich_querying_2012}.
Given a Markov process and uncertain locations of moving objects, \cite{emrich_querying_2012} proposed a modified matrix computation method to efficiently compute the probability of a user appearing in certain region during certain time period.  
However, the technique in \cite{emrich_querying_2012} can be only used to compute the probability of the presence of  a user in a region during a time period which we refer to as \textsc{Presence} event, 
whereas we extend the technique to a general set of spatiotemporal events. 


\section{Conclusion and Future Work}
In this paper, we investigate a new type of pivacy goal: protecting spatiotemporal event, which has not been studied in literature.
We  formally define spatiotemporal events and design a privacy metric extending the notion of differential privacy.
We proposed PriSTE, a framework integrating an LPPM for protecting the spatiotemporal event privacy.
 An interesting direction is to find optimal way for achieving both location privacy and spatiotemporal event privacy.
 Another question is how can we design a generic mechanism  for spatiotemporal event privacy without relying on an LPPM.



\vspace{-1pt}

\bibliographystyle{IEEEtran}
\bibliography{ref/stevent}

\begin{thebibliography}{10}
\providecommand{\url}[1]{#1}
\csname url@samestyle\endcsname
\providecommand{\newblock}{\relax}
\providecommand{\bibinfo}[2]{#2}
\providecommand{\BIBentrySTDinterwordspacing}{\spaceskip=0pt\relax}
\providecommand{\BIBentryALTinterwordstretchfactor}{4}
\providecommand{\BIBentryALTinterwordspacing}{\spaceskip=\fontdimen2\font plus
\BIBentryALTinterwordstretchfactor\fontdimen3\font minus
  \fontdimen4\font\relax}
\providecommand{\BIBforeignlanguage}[2]{{%
\expandafter\ifx\csname l@#1\endcsname\relax
\typeout{** WARNING: IEEEtran.bst: No hyphenation pattern has been}%
\typeout{** loaded for the language `#1'. Using the pattern for}%
\typeout{** the default language instead.}%
\else
\language=\csname l@#1\endcsname
\fi
#2}}
\providecommand{\BIBdecl}{\relax}
\BIBdecl

\bibitem{golle_anonymity_2009}
P.~Golle and K.~Partridge, ``\BIBforeignlanguage{en}{On the anonymity of
  {Home/Work} location pairs},'' in \emph{\BIBforeignlanguage{en}{Lecture Notes
  in Computer Science}}, 2009, pp. 390--397.

\bibitem{recabarren_what_2017}
R.~Recabarren and B.~Carbunar, ``What does the crowd say about you? evaluating
  aggregation-based location privacy,'' in \emph{{WPES}}, vol. 2017, 2017, pp.
  156--176.

\bibitem{argyros_evaluating_2017}
G.~Argyros, T.~Petsios, S.~Sivakorn, A.~D. Keromytis, and J.~Polakis,
  ``Evaluating the privacy guarantees of location proximity services,''
  \emph{{ACM} Trans. Priv. Secur.}, vol.~19, no.~4, pp. 12:1--12:31, 2017.

\bibitem{krumm_survey_2009}
J.~Krumm, ``\BIBforeignlanguage{en}{A survey of computational location
  privacy},'' \emph{\BIBforeignlanguage{en}{Personal and Ubiquitous
  Computing}}, vol.~13, no.~6, pp. 391--399, 2009.

\bibitem{wernke_classification_2014}
M.~Wernke, P.~Skvortsov, F.~D{\"u}rr, and K.~Rothermel,
  ``\BIBforeignlanguage{en}{A classification of location privacy attacks and
  approaches},'' \emph{\BIBforeignlanguage{en}{Personal and Ubiquitous
  Computing}}, vol.~18, no.~1, pp. 163--175, 2014.

\bibitem{chatzikokolakis_methods_2017}
K.~Chatzikokolakis, E.~{ElSalamouny}, C.~Palamidessi, and P.~Anna,
  ``\BIBforeignlanguage{English}{Methods for location privacy: A comparative
  overview},'' \emph{\BIBforeignlanguage{English}{Foundations and Trends® in
  Privacy and Security}}, vol.~1, no.~4, pp. 199--257, 2017.

\bibitem{liu_location_2018}
B.~Liu, W.~Zhou, T.~Zhu, L.~Gao, and Y.~Xiang, ``Location privacy and its
  applications,'' \emph{{IEEE} Access}, pp. 17\,606--17\,624, 2018.

\bibitem{andres_geo-indistinguishability:_2013}
M.~E. Andr{\'e}s, N.~E. Bordenabe, K.~Chatzikokolakis, and C.~Palamidessi,
  ``Geo-indistinguishability: differential privacy for location-based
  systems,'' in \emph{{CCS}}, 2013, pp. 901--914.

\bibitem{xiao_protecting_2015}
Y.~Xiao and L.~Xiong, ``Protecting locations with differential privacy under
  temporal correlations,'' in \emph{{CCS}}, 2015, pp. 1298--1309.

\bibitem{gruteser_anonymous_2003}
M.~Gruteser and D.~Grunwald, ``Anonymous usage of location-based services
  through spatial and temporal cloaking,'' in \emph{{MobiSys}}, 2003, pp.
  31--42.

\bibitem{dwork_differential_2008}
C.~Dwork, ``Differential privacy: A survey of results,'' in \emph{{TAMC}},
  2008, pp. 1--19.

\bibitem{chatzikokolakis_predictive_2014}
K.~Chatzikokolakis, C.~Palamidessi, and M.~Stronati,
  ``\BIBforeignlanguage{en}{A predictive differentially-private mechanism for
  mobility traces},'' in \emph{\BIBforeignlanguage{en}{Lecture Notes in
  Computer Science}}, 2014, no. 8555, pp. 21--41.

\bibitem{theodorakopoulos_prolonging_2014}
G.~Theodorakopoulos, R.~Shokri, C.~Troncoso, J.-P. Hubaux, and J.-Y. Le~Boudec,
  ``Prolonging the hide-and-seek game: Optimal trajectory privacy for
  location-based services,'' in \emph{{WPES}}, 2014, pp. 73--82.

\bibitem{dwork_differential_2010}
C.~Dwork, M.~Naor, T.~Pitassi, and G.~N. Rothblum, ``Differential privacy under
  continual observation,'' in \emph{STOC}, 2010, pp. 715--724.

\bibitem{schusterbockler_introduction_2007}
B.~{Schuster‐B{\"o}ckler} and A.~Bateman, ``\BIBforeignlanguage{en}{An
  introduction to hidden markov models},''
  \emph{\BIBforeignlanguage{en}{Current Protocols in Bioinformatics}}, vol.~18,
  no.~1, pp. A.3A.1--A.3A.9, 2007.

\bibitem{pardalos_quadratic_1991}
P.~M. Pardalos and S.~A. Vavasis, ``Quadratic programming with one negative
  eigenvalue is {NP-hard},'' \emph{Journal of Global Optimization}, vol.~1,
  no.~1, pp. 15--22, 1991.

\bibitem{dwork_pan-private_2010}
C.~Dwork, M.~Naor, T.~Pitassi, G.~N. Rothblum, and S.~Yekhanin, ``Pan-private
  streaming algorithms.'' in \emph{{ICS}}, 2010, pp. 66--80.

\bibitem{xiao_loclok:_2017}
Y.~Xiao, L.~Xiong, S.~Zhang, and Y.~Cao, ``{LocLok:} location cloaking with
  differential privacy via hidden markov model,'' \emph{{VLDB}}, vol.~10,
  no.~12, pp. 1901--1904, 2017.

\bibitem{zheng_geolife:_2010}
Y.~Zheng, X.~Xie, and W.-Y. Ma, ``{GeoLife:} a collaborative social networking
  service among user, location and trajectory,'' \emph{{IEEE} Data Eng. Bull.},
  vol.~33, no.~2, pp. 32--39, 2010.

\bibitem{ghinita_preventing_2009}
G.~Ghinita, M.~L. Damiani, C.~Silvestri, and E.~Bertino, ``Preventing
  velocity-based linkage attacks in location-aware applications,'' in
  \emph{{SIGSPATIAL}}, 2009, pp. 246--255.

\bibitem{ardagna_obfuscation-based_2011}
C.~A. Ardagna, M.~Cremonini, S.~D. C.~d. Vimercati, and P.~Samarati, ``An
  obfuscation-based approach for protecting location privacy,'' \emph{{IEEE}
  TDSC}, vol.~8, no.~1, pp. 13--27, 2011.

\bibitem{hwang_novel_2012}
R.~H. Hwang, Y.~L. Hsueh, and H.~W. Chung, ``A novel time-obfuscated algorithm
  for trajectory privacy protection,'' pp. 126--139, 2014.

\bibitem{rodriguez-carrion_entropy-based_2015}
A.~Rodriguez-Carrion, D.~Rebollo-Monedero, J.~Forn{\'e}, C.~Campo,
  C.~Garcia-Rubio, J.~Parra-Arnau, and S.~K. Das,
  ``\BIBforeignlanguage{en}{Entropy-based privacy against profiling of user
  mobility},'' \emph{\BIBforeignlanguage{en}{Entropy}}, vol.~17, no.~6, pp.
  3913--3946, 2015.

\bibitem{shokri_quantifying_2011}
R.~Shokri, G.~Theodorakopoulos, J.-Y. Le~Boudec, and J.-P. Hubaux,
  ``Quantifying location privacy,'' in \emph{{SP}}, 2011, pp. 247--262.

\bibitem{bordenabe_optimal_2014}
N.~E. Bordenabe, K.~Chatzikokolakis, and C.~Palamidessi, ``Optimal
  geo-indistinguishable mechanisms for location privacy,'' in \emph{{CCS}},
  2014, pp. 251--262.

\bibitem{shokri_privacy_2015}
R.~Shokri, ``Privacy games: Optimal user-centric data obfuscation,''
  \emph{{PET}}, vol. 2015, no.~2, 2015.

\bibitem{xu_trajectory_2017}
F.~Xu, Z.~Tu, Y.~Li, P.~Zhang, X.~Fu, and D.~Jin, ``Trajectory recovery from
  ash: User privacy is {NOT} preserved in aggregated mobility data,'' in
  \emph{{WWW}}, 2017, pp. 1241--1250.

\bibitem{cao_quantifying_2017}
Y.~Cao, M.~Yoshikawa, Y.~Xiao, and L.~Xiong, ``Quantifying differential privacy
  under temporal correlations,'' in \emph{{ICDE)}}, 2017, pp. 821--832.

\bibitem{liao_learning_2004}
L.~Liao, D.~Fox, and H.~Kautz, ``Learning and inferring transportation
  routines,'' in \emph{{AAAI}}, 2004, pp. 348--353.

\bibitem{qiao_putmode:_2010}
S.~Qiao, C.~Tang, H.~Jin, T.~Long, S.~Dai, Y.~Ku, and M.~Chau,
  ``\BIBforeignlanguage{en}{{PutMode:} prediction of uncertain trajectories in
  moving objects databases},'' \emph{\BIBforeignlanguage{en}{Applied
  Intelligence}}, vol.~33, no.~3, pp. 370--386, 2010.

\bibitem{li_swarm:_2010}
Z.~Li, B.~Ding, J.~Han, and R.~Kays, ``Swarm: Mining relaxed temporal moving
  object clusters,'' \emph{{VLDB}}, vol.~3, no. 1-2, pp. 723--734, 2010.

\bibitem{emrich_querying_2012}
T.~Emrich, H.~P. Kriegel, N.~Mamoulis, M.~Renz, and A.~Zufle, ``Querying
  uncertain spatio-temporal data,'' in \emph{ICDE}, 2012, pp. 354--365.

\end{thebibliography}
\newpage

\newpage

\begin{appendices}
	
\section{Proofs}

\subsection{Proof of Lemma \ref{lemma-post-before}}
\begin{proof}
	At timestamps $1$, {\footnotesize $\boldsymbol\alpha_1=[\boldsymbol\pi, \textbf{0}]\circ\tilde{\textbf{p}}_{o_1}$}. 
	For $t>1$, 
	{\footnotesize $\boldsymbol\alpha_t=\boldsymbol\alpha_1\textbf{M}_{1}\tilde{\textbf{p}}_{o_2}^\textbf{D}\cdots\textbf{M}_{t-1}\tilde{\textbf{p}}_{o_t}^\textbf{D}	=[\boldsymbol\pi,\textbf{0}]\left(  \tilde{\textbf{p}}_{o_1}^\textbf{D}\prod_{i=2}^{t}(\textbf{M}_{i-1}\tilde{\textbf{p}}_{o_i}^\textbf{D})\right)$}. By Lemma \ref{theo-prior}, 
	{\footnotesize $\Pr(\textsc{Event},o_1,o_2,\cdots,o_t)=\boldsymbol\alpha_t\prod_{i=t}^{end-1}\textbf{M}_i[\textbf{0},\textbf{1}]^\intercal$}. 
	Then Equation (\ref{eqn-post-before}) can be derived. 
\end{proof}

\subsection{Proof of Lemma \ref{lemma-post-after}}
\begin{proof}
	{\small $\Pr(o_1,o_2,\cdots,o_t,\textsc{Event}\textrm{ is true})=\sum_{k} \Pr \big(u_{end}=k,o_1,o_2,$}  {\small$\cdots,o_t, \textsc{Event}\textrm{ is true} \big)$}.
	By forward-backward algorithm, {\small $\Pr(u^{t}=s_{k},o_1,o_2,\cdots,o_t)=\alpha_{t}^{k}\beta_{t}^{k}$}. Thus we only need to derive the $\boldsymbol\alpha_{end}$ and $\boldsymbol\beta_{end}$ and compute the sum of {\small $\boldsymbol\alpha_{end}\circ \boldsymbol\beta_{end}$} in the world where the $\textsc{Event}$ is true.  
	By Lemma \ref{lemma-post-before}, 
	{\small $\boldsymbol\alpha_{end}=[\boldsymbol\pi,\textbf{0}]\left(  \tilde{\textbf{p}}_{o_1}^\textbf{D}\prod_{i=2}^{end}(\textbf{M}_{i-1}\tilde{\textbf{p}}_{o_i}^\textbf{D})\right)$}. 
	By backward algorithm, {\small $\boldsymbol\beta_{end}
		=[\textbf{1},\textbf{1} ]\prod_{i=t-1}^{end}(\tilde{\textbf{p}}_{o_{i+1}}^\textbf{D}\textbf{M}_i^\intercal)$}
	with {\small $\boldsymbol\beta_t=[\textbf{1},\textbf{1} ]$}. 
	Thus {\small $\Pr(\textsc{Event},o_1,o_2,\cdots,o_t)=(\boldsymbol\alpha_{end}\circ\boldsymbol\beta_{end})[\textbf{0},\textbf{1}]^\intercal
		=\boldsymbol\alpha_{end}(\boldsymbol\beta_{end} \circ [\textbf{0},\textbf{1}])^\intercal$}, which is equal to Equation (\ref{eqn-post-after}).
\end{proof}

\subsection{Proof of Theorem \ref{theo-eps-delta-DP}}

\begin{proof}[Proof Sketch]
	By Definition \ref{def-eps-delta-DP}, it is equivalent to prove {\small $f_{1}(\boldsymbol\pi)\leq 0$} and {\small $f_{2}(\boldsymbol\pi)\leq 0$} where 
	{\small $f_{1}(\boldsymbol\pi)=\frac{\Pr(o_{1},o_{2},\cdots,o_{T}, \textsc{Event})}{\Pr(\textsc{Event})}-e^{\epsilon}\frac{\Pr(o_{1},o_{2},\cdots,o_{T}, \lnot\textsc{Event})}{\Pr(\lnot\textsc{Event})}$} 
	and 
	{\small $f_{2}(\boldsymbol\pi)=\frac{\Pr(o_{1},o_{2},\cdots,o_{T}, \lnot\textsc{Event})}{\Pr(\lnot\textsc{Event})}-e^{\epsilon}\frac{\Pr(o_{1},o_{2},\cdots,o_{T}, \textsc{Event})}{\Pr(\textsc{Event})}$}. By Lemma \ref{theo-prior} and \ref{lemma-post-after}, Theorem \ref{theo-eps-delta-DP} can be derived. 
\end{proof}

\section{Naive Solutions}

\subsection{Computing Prior Probability of an Event}

Considering an event is  a set of Boolean expression of (location, time) predicates combined with AND and OR, a naive approach would be to enumerate all possible cases for the event and sum (correspond to OR) the product (correspond to AND) of the probabilities of each location predicate and such an approach would require exponential computation time. 
Due to space limitation, we omit a detailed algorithm for this naive solution.
Instead, we show an example as below.
\begin{figure}[h]
	\centering
	\includegraphics[width=7cm]{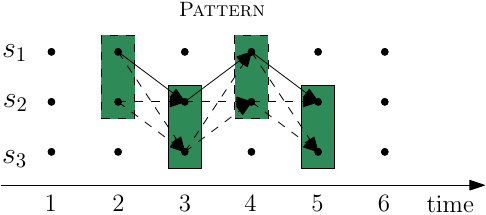}
	\caption{{\footnotesize $\Pr(\textsc{Pattern})=\sum Pr(\textrm{all trajectories rendering \textsc{Pattern}})$ for the $2^{4}$ trajectories satisfying the \textsc{Pattern} event defined above.}}
	\label{Figure-example-naive}
\end{figure}
\begin{example}
	\label{example-prior-pattern}
	In Fig.\ref{Figure-example-naive}, a \textsc{Pattern} event is defined by the shaded regions during timestamp $2$ to $5$. Thus $start=2$, $end=5$. The regions are {\small $\textbf{s}_2=[1,1,0]^{\intercal}$}, {\small $\textbf{s}_3=[0,1,1]^{\intercal}$}, {\small $\textbf{s}_4=[1,1,0]^{\intercal}$}, {\small $\textbf{s}_5=[0,1,1]^{\intercal}$}. To derive the probability of one trajectory, e.g. the solid lines in Fig.\ref{Figure-example-naive}, it would be 
	{\small $\Pr(u^{2}=s_{1})*
		\Pr(u^{3}=s_{2}|u^{2}=s_{1})*\Pr(u^{4}=s_{1}|u^{3}=s_{2},u^{2}=s_{1})*\Pr(u^{5}=s_{1}|u^{4}=s_{1},u^{3}=s_{2},u^{2}=s_{1})$}. Because there are $2^{4}$ trajectories for the \textsc{Pattern} event, {the prior probability of  \textsc{Pattern}, i.e.,} $ \Pr $(\textsc{Pattern}) is the sum of $2^{4}$ such probabilities.
\end{example}

\subsection{Computing Join Probability of an Event}

\begin{algorithm}[h]
	\scriptsize
	\caption{Naive Algorithm to Derive the Joint Probability of a \textsc{Pattern}}
	\begin{algorithmic}
		\Require{
			$\textbf{M}$, $\textbf{p}_{start-1}$: the probability at timestamp $start-1$, $\tilde{\textbf{p}}_{o_t}$, $\textsc{Pattern}$ in timestamp $start,\cdots, end$
		}
		\State{$p_{pattern}\gets 0$}
		\For{traj in \textsc{Pattern}}
		\Comment{exponential trajectories}
		\State{$\textbf{p}_{traj}\gets (\textbf{p}_{start-1}\textbf{M})\circ \tilde{\textbf{p}}_{o_t}$}
		\State{$p_{traj}=\textbf{p}_{traj}[u_{start}\ in\ traj]$}
		\For{$t$ in $\{start+1,\cdots,end\}$}
		\State{$s_{t-1}\gets u_{t-1}\ in\ traj$}
		\State{$s_{t}\gets u^{t}\ in\ traj$}
		\State{$p_{traj}\gets p_{traj}*m_{s_{t-1}s_{t}}*\tilde{\textbf{p}}_{o_t}[{s_{t}}]$}
		\EndFor
		\State{$p_{pattern}\gets p_{pattern}+p_{traj}$}
		\EndFor
	\end{algorithmic}
	\label{alg-naive}
\end{algorithm}

Let traj be a trajectory of the \textsc{Pattern}. There will be $|traj|$ trajectories of the \textsc{Pattern}. For example, in Figure \ref{Figure-example-naive}, the solid line is $traj=\{u^{2}=s_{1},u^{3}=s_{2},u^{4}=s_{1},u^{5}=s_{2}\}$, which means $u^{2}$ in traj is $s_{1}$, $\cdots$, $u^{5}$ in traj is $s_{2}$. Let $\textbf{M}$ be the transition matrix where $m_{ij}$ is the transition probability from state $i$ to state $j$. 
$\tilde{\textbf{p}}_{o_t}$ denotes the emission probability of observing $o_t$, 
$\tilde{\textbf{p}}_{o_t}=[
\Pr(o_t|u^t=s_1), \Pr(o_t|u^t=s_2),\cdots,\Pr(o_t|u^t=s_m)]$. Thus $\tilde{\textbf{p}}_{o_t}[s_{1}]=\Pr(o_t|u^t=s_1)$, $\cdots$, $\tilde{\textbf{p}}_{o_t}[s_{m}]=\Pr(o_t|u^t=s_m)$.
Let $\Pr(o_{start},\cdots,o_{end},traj)$ be the joint probability of the observations and the trajectory. Then $\Pr(traj | o_{start},\cdots,o_{end})=\frac{\Pr(o_{start},\cdots,o_{end},traj)}{\Pr(o_{start},\cdots,o_{end})}$. Thus we focus on the joint probability $\Pr(o_{start},\cdots,o_{end},traj)$.

\noindent{\bf Setup.} Let $start=2$, $\textbf{p}_{start-1}=\boldsymbol\pi$. For a given $\boldsymbol\pi$, Algorithm \ref{alg-naive} can be used to derive the joint probability of a \textsc{Pattern}. The runtime can be compared with the runtime of Equation (\ref{eqn-post-before}), which is also the joint probability. 

Note that to derive Equation (\ref{eqn-post-before}), we prefer ``vector*matrix'' instead of ``matrix*matrix'' for efficiency. Calculate it from left to right so that no matrix multiplication should be conducted.

\section{Examples}

\subsection{An Example of Prior Probability Computation}
We use an example to describe the  details in computation. 
\begin{example}[Prior Probability Computation]
	\label{example-running-2}
	Let us consider the computation of Example \ref{example-prior-presence}. For the $\textsc{Presence}$ event defined at $t=3$ and $t=4$, the transition matrix at $t=2$ and $t=3$ derived by Equation (\ref{new-M-1}) is the left matrix below; while the transition matrix at $t=1$ and $t\geq 4$ derived by Equation (\ref{new-M-2}) is the right matrix below. 
	\begin{myAlignSSS}
		\hspace{-1mm}
		\left[
		\begin{array}{cccccc}
			0 &0&0.7&0.1&0.2&0\\
			0&0&0.5 & 0.4&0.1&0\\
			0&0&0.9&0&0.1&0\\
			0&0&0&0.1&0.2&0.7\\
			0&0&0&0.4&0.1&0.5\\
			0&0&0&0&0.1&0.9
		\end{array}
		\right]
		\left[
		\begin{array}{cccccc}
			0.1 &0.2&0.7&0&0&0\\
			0.4&0.1&0.5 & 0&0&0\\
			0&0.1&0.9&0&0&0\\
			0&0&0&0.1&0.2&0.7\\
			0&0&0&0.4&0.1&0.5\\
			0&0&0&0&0.1&0.9
		\end{array}
		\right]
	\end{myAlignSSS}
	
	\vspace{-10pt}
	\begin{minipage}{0.48\textwidth}
		\begin{tabular}{p{23pt}p{110pt}p{105pt}}
			&$\textbf{M}_{2}, \textbf{M}_{3}$ & $\textbf{M}_{1}, \textbf{M}_{4}, \textbf{M}_{5}$
		\end{tabular}
	\end{minipage}
	
	Thus by Lemma \ref{theo-prior}, {\footnotesize $\Pr(\textsc{Presence})=[\boldsymbol\pi,\textbf{0}]\textbf{M}_{1}\textbf{M}_{2}\textbf{M}_{3}[\textbf{0},\textbf{1}]^{\intercal}$}, which is a function of $\boldsymbol\pi$ as {\footnotesize $\Pr(\textsc{Presence})=\boldsymbol\pi[0.28, 0.298,0.226]^{\intercal}$}.
\end{example}

\end{appendices}

\end{document}